\documentclass{article}
\usepackage{authblk}
\usepackage[utf8]{inputenc}
\usepackage{parskip}
\usepackage{tikz}
\usepackage{blindtext}
\usepackage{multicol}
\usepackage{textcomp}
\usepackage{comment}
\usepackage[style=numeric-comp, sorting=none]{biblatex}
\usepackage{graphicx}
\usepackage{epsfig}
\usepackage{amsmath}
\usepackage{subcaption}
\graphicspath{ {./image/} }
\usepackage{amssymb}
\usepackage{float}
\usepackage{appendix}
\usepackage{hyperref}
\hypersetup{
    colorlinks=true,
    linkcolor=blue,
    filecolor=magenta,      
    urlcolor=cyan,
    pdftitle={Overleaf Example},
    pdfpagemode=FullScreen,
    }
\usepackage{geometry}
 \usepackage{soul}

\begin{document}
\date{}
\title{Quasinormal modes of Einstein--scalar--Gauss--Bonnet black holes}
\author[1]{Prosenjit Paul\thanks{prosenjitpaul629@gmail.com}}
\affil[1]{Indian Institute Of Engineering Science and Technology (IIEST), Shibpur-711103, WB, India}
\maketitle
\begin{abstract}
In this paper, we investigate quasinormal modes of scalar and electromagnetic 
fields in the background of Einstein--scalar--Gauss--Bonnet (EsGB) black holes. Using 
the scalar and electromagnetic field equations in the vicinity of the EsGB black hole, 
we study nature of the effective potentials. The dependence of real and imaginary parts of the 
fundamental quasinormal modes on parameter $p$ (which is related to the Gauss--Bonnet coupling parameter $\alpha$) for different values of multipole numbers $l$ are 
studied. We analyzed the effects of massive scalar fields on the EsGB black hole, which 
tells us the existence of quasi--resonances. In the eikonal regime, we find the 
analytical expression for the quasinormal frequency and show that the correspondence between the eikonal quasinormal modes and null geodesics is valid in the EsGB theory for the test fields. Finally, we study grey-body 
factors of the electromagnetic fields for different multipole numbers $l$, which 
deviates from Schwarzschild's black hole.   
\end{abstract}

\section{Introduction}\label{sec:1}
General Relativity (GR) is the simplest theory of gravitation, which
is consistent with various astrophysical observations, like gravitational waves, 
black holes, etc. Nevertheless, GR opens some challenging problems. These are the 
existence of singularities at the center of the black holes, a complete theory
of quantum gravity, the problem of dark matter/energy, cosmic inflation, and others. To solve 
these problems theorists introduce alternative approaches to gravity. There are a number of such approaches to gravity available in the literature, e.g.,
string--theory--inspired gravities, adding higher-order terms in the curvature tensor
to the Einstein–-Hilbert action of GR.

One of the well-motivated alternative theories of gravity is the EsGB theory. In this theory, the scalar field is nonminimally coupled to the Gauss--Bonnet (GB) term.
The EsGB theories also arise in the low energy limit of string theory \cite{Metsaev:1987zx}.
The lowest order correction to the Einstein-Hilbert action is the GB term, which 
is quadratic in curvature tensor, but this theory has pure divergence in $4D$ 
spacetime. Alternatively, in recent years Glavan and Lin \cite{Glavan:2019inb} have removed the divergence in $4D$ GB gravity by replacing the GB parameter $\alpha$ to $\alpha/D-4$, where $D$ is the number of spacetime dimensions. However, later it was shown that this naive regulariation scheme does not lead to the well-defined theory of gravity. Nevertheless, the black hole solutions \cite{Fernandes:2020rpa,Konoplya:2020ibi} obtained as results of such regularazation proved out to be also solutions in the well-defined theories \cite{Aoki:2020lig}. On the contrary, the EsGB theories, we are interested here, are free of such kind of problems, because in EsGB theories the divergence is absent due to the coupling between the GB term and scalar fields.

Isolated black holes are very simple objects and can be described by the three 
parameters: mass, charge, and spin. However, the actual situation differs from this. A black 
hole at the center of galaxies is surrounded by matter distribution, such as 
an accretion disk, jets and outflows, stars, etc. Therefore, a black hole interacts 
with its surroundings matter distribution. Even in the absence of any matter 
distribution around black hole, it will interact with the vacuum and 
produce a pair of particles and anti-particle, which is known as Hawking 
radiation. Therefore, a  dynamical, interacting black hole can not be described by mass, charge, and 
spin only, and one needs to consider the perturbations theory of black holes. There 
has been a growing interest in the perturbation theory of black holes \cite{Kokkotas:1999bd,Berti:2009kk,Konoplya:2011qq}. 
There are a number of reasons for interest in proper frequencies of such out-of-equilibrium black holes, called {\it quasinormal modes.} First of all, the LIGO and VIRGO scientific 
collaborations detect gravitational-wave signals from black holes \cite{LIGOScientific:2016aoc} 
and it is consistent with Einstein's theories of gravity, though, due to the large uncertainty in the determining of the spin and mass of the black hole, the large window for alternative theories remains \cite{Konoplya:2016pmh,Yunes:2016jcc}. The 
dominating influence on such a signal is expected to come from the 
quasinormal modes exhibiting the lowest frequency, 
referred to as the fundamental mode.

Various black hole solutions in higher-order theories of gravity, its quasinormal 
modes, and Hawking radiation are studied extensively. The EsGB gravity was studied in Refs. \cite{Kanti:1995vq,Antoniou:2017hxj,Collodel:2019kkx}, where the most important black hole solutions were found numerically. 
The analytical black hole solution of EsGB theories using continued fraction approximation (CFA) and its shadow 
was studied in Ref. \cite{Konoplya:2019fpy}. The gravitational quasinormal modes of Einstein--dilaton--Gauss--Bonnet (EdGB)
solutions are studied in Refs. \cite{Blazquez-Salcedo:2017txk,Bryant:2021xdh}.
The spontaneous scalarization, black hole sensitivities, and linear stability of the EsGB 
black hole are investigated in Refs. \cite{East:2021bqk,Julie:2022huo,Minamitsuji:2023uyb}.
Black hole solutions in EdGB, Einstein--Wely, and Einstein cubic gravity investigated in Refs. \cite{Kleihaus:2015aje,Kokkotas:2017ymc,Kokkotas:2017zwt,Hennigar:2018hza}, 
while its quasinormal modes and Hawking radiation are studied in Refs. \cite{Zinhailo:2018ska,Konoplya:2019hml,Konoplya:2019ppy,Zinhailo:2019rwd,Konoplya:2020jgt}.
In addition, quasinormal modes of Kaluza-Klein-like black holes in the Einstein-Gauss-Bonnet theory were considered in \cite{Cao:2023zhy}. Furthermore, quasinormal modes of string-corrected $d$-dimensional black holes and noncommutative Schwarzschild black holes are analysed in Refs. \cite{Cuyubamba:2016cug,Yoshida:2015vua,Moura:2021eln,Moura:2021nuh,Moura:2022gqm,Campos:2021sff}.
The gravitational perturbations of numerically obtained EdGB black holes
are studied in Ref. \cite{Blazquez-Salcedo:2017txk}. In the eikonal 
limit, the quasinormal modes for gravitational perturbations of 
EsGB black holes were obtained in Ref. \cite{Bryant:2021xdh}. However, the scalar and 
electromagnetic perturbations either of analytically or numerically obtained EsGB black hole solution are not analyzed, which provides us an opportunity to fill this gap. The scalar field will be studied not only in the massless limit, but also for the non-zero massive term. The latter has a number of motivations, because an effective mass term appears in the wave equation as a result of introduction of extra dimensions \cite{Seahra:2004fg,Ishihara:2008re}, and magnetic fields \cite{Konoplya:2008hj,Chen:2011jgd}. After all, massive long-lived modes and oscillatory tails \cite{Konoplya:2023fmh} may contribute into the very long gravitational waves observed recently via the Time Pulsar Array  \cite{NANOGrav:2023gor,NANOGrav:2023hvm}.

The paper is organized as follows. In section \ref{sec:2}, we outline the 
basics of EsGB theories in four dimensions and discuss the metric functions for black holes 
in EsGB gravity. The quasinormal modes for massless/massive scalar fields and 
electromagnetic fields are studied in section \ref{sec:3} for different scalar coupling 
functions. Furthermore, we derived the analytical formula for quasinormal frequency in the eikonal regime. 
In section \ref{sec:4}, we discuss the grey--body factors for electromagnetic fields.
Finally, we summarize our results in the conclusions section.

\section{Einstein--scalar--Gauss-Bonnet Black Holes}\label{sec:2}
The action for EsGB theories in $4D$ can be written as 
\begin{equation}\label{eq:1}
 I= \int dx^4 \sqrt{-g}  \biggr[ \frac{R}{\kappa^2} + \alpha f(\varphi) R_{GB}^2 - \frac{1}{2} \nabla_{\mu} \varphi \nabla^{\mu} \varphi   \biggr],  
\end{equation}
where $g$ is  determinant of the metric $g_{\rho \sigma}$, we take $\kappa=16 \pi G c^{-4}=1$, $\alpha$ is the GB coupling constant and $R_{GB}^2$ is defined as 
\begin{equation}\label{eq:2}
R_{GB}^2=R_{\mu \nu \rho \sigma} R^{\mu \nu \rho \sigma} -4 R_{\mu \nu } R^{\mu \nu } + R^2,
\end{equation}
and $f(\varphi)$ is the arbitrary smooth function of the scalar field $\varphi$, which is known as GB coupling functional. The metric of static and spherically symmetric black holes in EsGB theory can be written in the following form \cite{Konoplya:2019fpy}
\begin{equation}\label{eq:3}
    ds^2 = -g_{tt} dt^2 + g_{rr} dr^2 + r^2 \Bigl(d\theta^2 + \sin^2{\theta} d\phi^2 \Bigl),
\end{equation}
where the analytically approximated metric functions were found in 
Ref. \cite{Konoplya:2019fpy}. The functions $g_{tt}$ and $g_{rr}$ are 
written up to fourth order for different GB coupling functional in 
the Appendix \ref{sec:6}. Here we consider quadratic, cubic, quartic, inverse and
Logarithmic Gauss-Bonnet coupling functional. 

To parameterize the family of EsGB black holes solution, we will introduce the 
dimension--less parameter p as
\begin{equation}\label{eq:4}
    p \equiv \frac{96 \alpha^2 f^{\prime}(\varphi)^2 }{r_{0}^4}, \\ 0 \leq p <1,
\end{equation}
where $r_{0}$ is the position of the event horizon, $\varphi_{0}$ is the 
scalar fields at the event horizon. In the limit $p=0$, one can obtain 
Schwarzschild black hole.

\section{Quasinormal Modes of Test Fields}\label{sec:3}
In this paper, we consider quasinormal modes of test fields in the background 
of EsGB black holes. The equation for scalar, and electromagnetic in the 
background of the EsGB black hole is given by
\begin{subequations}\label{eq:5}
\begin{align}
\frac{1}{\sqrt{-g}} \partial_{\mu} \Bigl( \sqrt{-g} g^{\mu \nu} \partial_{\nu} \Phi\Bigl) -m^2 \Phi&=0, \label{eq:5a} \\ 
\frac{1}{\sqrt{-g}} \partial_{\mu} \Bigl( \sqrt{-g} g^{\sigma \mu} g^{\rho \nu} F_{\rho \sigma}\Bigl) &=0, \label{eq:5b} 
\end{align}
\end{subequations}
where $F_{\rho \sigma} = \partial_{\rho} A_{\sigma} - \partial_{\sigma} A_{\rho}$ and $A_{\mu}$ is the vector potential. To do the separation of variables we will introduce the radial function $R_{\omega l}(r)$ and spherical harmonics $Y_{l}(\theta, \phi)$ as

\begin{equation}\label{eq:6}
\Phi(t,r,\theta,\phi)= e^{\pm i \omega t} R_{\omega l}(r) Y_{l}(\theta, \phi), 
\end{equation}
where $l$ is the angular number. After the separation of variables equation \eqref{eq:5} can be written in the following Schrodinger-like wave equation \cite{Konoplya:2006rv,Konoplya:2019hml}
\begin{equation}\label{eq:7}
    \frac{d^2 \Psi}{dr_{*}^2} + \Bigl(\omega^2 - V_{i}(r)\Bigl) \Psi=0,
\end{equation}

where $r_{*}$ is the ``\emph{tortoise coordinate}", defined as
\begin{equation}\label{eq:8}
    dr_{*}= dr \sqrt{\frac{g_{rr}}{g_{tt}}}.
\end{equation}
The effective potentials of scalar $(i = s)$ and electromagnetic $(i = e)$ fields in the background of EsGB black hole are given by
\begin{subequations}\label{eq:9}
\begin{align}
V_{s} &= \frac{g_{rr} g_{tt}^{\prime} - g_{tt} g_{rr}^{\prime}}{2r g_{rr}^2} + g_{tt} \frac{l(l+1) }{r^2} +g_{tt} m, \label{eq:10a} \\ 
V_{e} &=  g_{tt} \frac{l(l+1) }{r^2}.   \label{eq:10b}
\end{align}
\end{subequations}

\begin{figure}[H]
    \centering
    \begin{subfigure}[b]{0.3\textwidth}
        \centering
        \includegraphics[width=\textwidth]{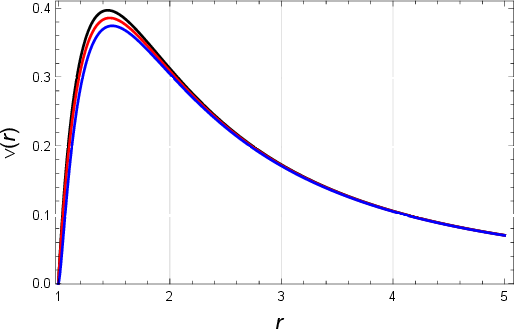}
        \caption{$f(\varphi)=\varphi^3$: $l=1$ \& $s=0$}
        \label{fig:1(e)}
    \end{subfigure}
    \hfill
    \begin{subfigure}[b]{0.3\textwidth}
        \centering
        \includegraphics[width=\textwidth]{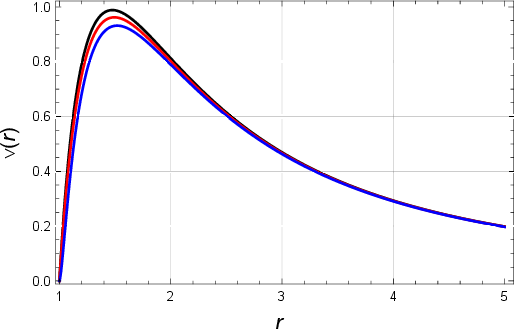}
        \caption{$f(\varphi)=\varphi^3$: $l=2$ \& $s=0$}
        \label{fig:1(f)}
    \end{subfigure}
     \hfill
    \begin{subfigure}[b]{0.3\textwidth}
        \centering
        \includegraphics[width=\textwidth]{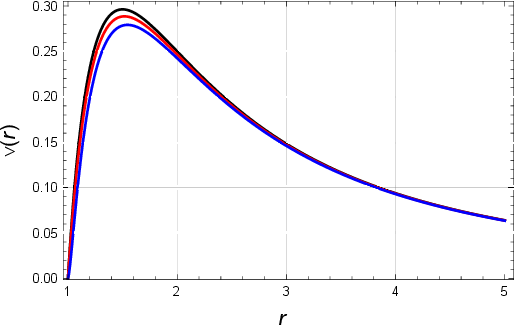}
        \caption{$f(\varphi)=\varphi^3$: $l=1$ \& $s=1$}
        \label{fig:1(g)}
    \end{subfigure}
     \hfill
    \begin{subfigure}[b]{0.3\textwidth}
        \centering
        \includegraphics[width=\textwidth]{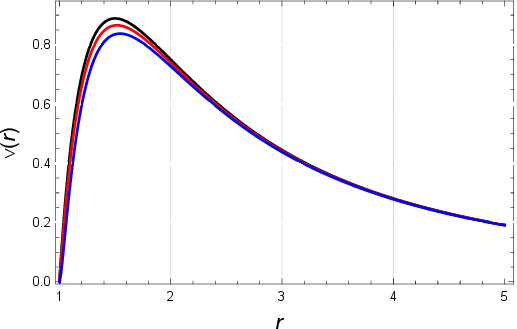}
        \caption{$f(\varphi)=\varphi^3$: $l=2$ \& $s=1$}
        \label{fig:1(h)}
    \end{subfigure}
\hfill
\begin{subfigure}[b]{0.3\textwidth}
        \centering
        \includegraphics[width=\textwidth]{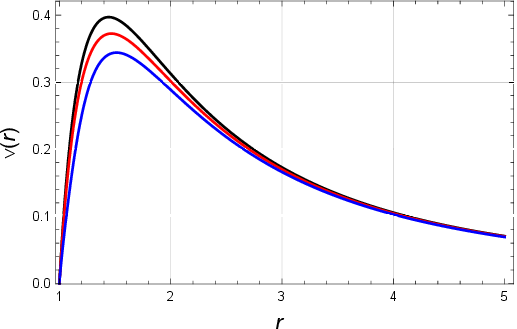}
        \caption{$f(\varphi)=\ln(\varphi)$: $l=1$ \& $s=0$}
        \label{fig:1(q)}
\end{subfigure}
\hfill
\begin{subfigure}[b]{0.3\textwidth}
        \centering
        \includegraphics[width=\textwidth]{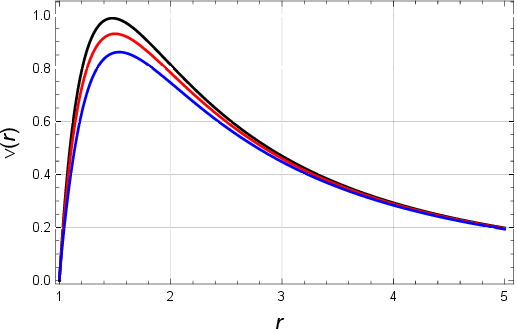}
        \caption{$f(\varphi)=\ln(\varphi)$: $l=2$ \& $s=0$}
        \label{fig:1(r)}
\end{subfigure}
\hfill
\begin{subfigure}[b]{0.3\textwidth}
        \centering
        \includegraphics[width=\textwidth]{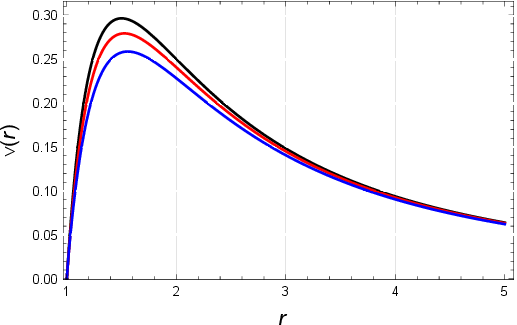}
        \caption{$f(\varphi)=\ln(\varphi)$: $l=1$ \& $s=1$}
        \label{fig:1(s)}
\end{subfigure}
\hfill
\begin{subfigure}[b]{0.3\textwidth}
        \centering
        \includegraphics[width=\textwidth]{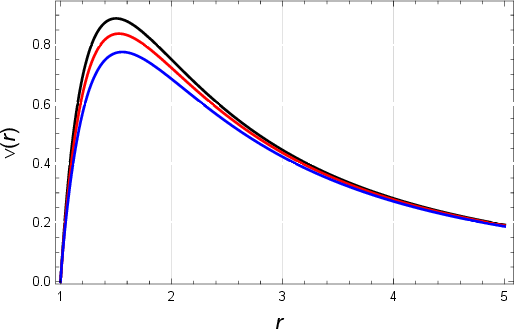}
        \caption{$f(\varphi)=\ln(\varphi)$: $l=2$ \& $s=1$}
        \label{fig:1(t)}
\end{subfigure}
\caption{Effective potential with $r_0=1$, black line denotes 
    $p=0.0$, red line denotes $p=0.4$ and blue lines denotes $p=0.8$}
\label{fig:1}
\end{figure}

\begin{figure}[H]
    \centering
    \begin{subfigure}[b]{0.3\textwidth}
        \centering
        \includegraphics[width=\textwidth]{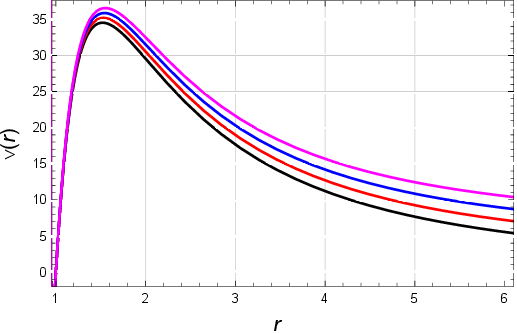}
        \caption{$f(\varphi)=\varphi^3$: $l=15$ \& $s=0$}
        \label{fig:2(b)}
    \end{subfigure}
    \hfill
    \begin{subfigure}[b]{0.3\textwidth}
        \centering
        \includegraphics[width=\textwidth]{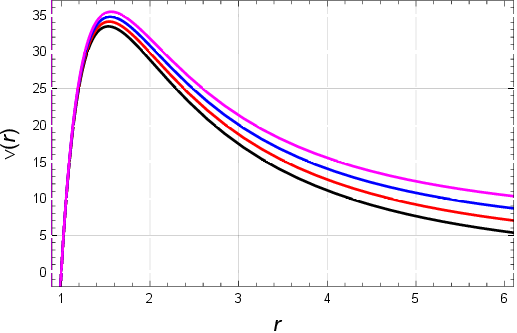}
        \caption{$f(\varphi)=\varphi^{-1}$: $l=15$ \& $s=1$}
        \label{fig:2(d)}
    \end{subfigure}
    \hfill
    \begin{subfigure}[b]{0.3\textwidth}
        \centering
        \includegraphics[width=\textwidth]{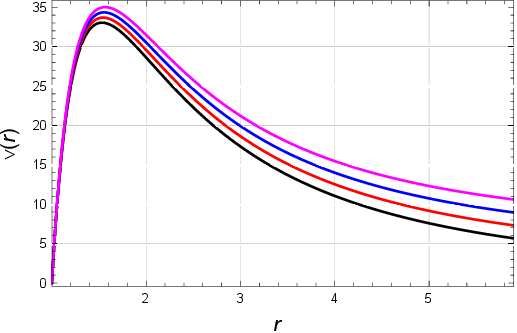}
        \caption{$f(\varphi)=\ln(\varphi)$: $l=15$ \& $s=0$}
        \label{fig:2(e)}
    \end{subfigure}
    \caption{Effective potential for massive scalar fields with $r_0=1$, $p=0.5$, black line denotes 
    $m=0.0$, red line denotes $m=2.0$, blue lines denotes $m=4.0$ and magenta line denotes
    $m=6.0$.}
    \label{fig:2}
\end{figure}

The dependence of effective potential on various parameters (e.g. $p$ and $l$)
for massless scalar fields and electromagnetic fields are shown in Fig. \ref{fig:1} for cubic and logarithmic coupling functional.
To express the radial coordinates in units of event horizon radius, we set
$r_{0}=1$. The form of effective potential is positive definite and it 
diminishes at both the event horizon and infinity. For massless scalar and 
electromagnetic fields as we increase the parameter $p$ height of 
the potential barrier decreases and the barrier height is maximum at 
$p=0$. It can be seen that as we increase the angular number $l$ 
height of the potential barrier becomes higher. For massive scalar fields,
the height of the potential barrier (Fig. \ref{fig:2}) increases with mass and the height of the potential barrier is minimum 
for $m=0$. The effective potential for others coupling functional exhibits similar behaviour.

The master equation \eqref{eq:7} can be solved using WKB approximation \cite{Schutz:1985km}.
The wave function $\Psi$ satisfies the following boundary condition
\begin{equation}\label{eq:10}
    \Psi \sim  e^{\pm i \omega r_{*}},
\end{equation}
at $r_{*} \to \pm \infty$, i.e. there are only incoming waves at the
event horizon and purely outgoing waves at spatial infinity. In the
asymptotic region, using the boundary condition one can solve the 
master equation using the WKB approximations up to the required order.
Expanding the potential into the Taylor series and using the master
equation, the solution of differential equation \eqref{eq:7} is obtained 
near the peak of the potential. Matching these two solutions the 
quasinormal frequency $\omega$ is derived. Finally, the frequency can be expressed 
as $\omega = \text{Re}(\omega) + i \text{Im}(\omega)$, where $\text{Re}(\omega)$ represent real oscillation 
frequency and $\text{Im}(\omega)$ represent damping rate. The quasinormal frequency
upto 6-th order WKB method has the following form
\begin{equation}\label{eq:11}
   \frac{i (\omega^2 - V_{0})}{\sqrt{-2 V_{0}^{\prime \prime}}} 
   - \sum_{i=2}^{i=6} \Lambda_{i} =n+\frac{1}{2},  
\end{equation}

where $\Lambda_{i}$ are the correction term up to sixth-order \cite{Schutz:1985km,Iyer:1986np,Konoplya:2003ii,Konoplya:2004ip}.
$n=0$, $1$, $2$ is the overtone number. To compute the quasinormal modes 
of massless scalar and vector fields we will use  sixth-order WKB formula with Pad\'e approximation \cite{Matyjasek:2017psv,Konoplya:2019hlu}
 $\Tilde{m}=5$.

It is worth of mentioning that special attention will be devoted to the $l\gg 1$, eikonal, regime, first of all because there is a correspondence between the eikonal quasinormal modes and some parameters of the null geodesics suggested in  \cite{Cardoso:2008bp} and further constrained in  \cite{Konoplya:2022gjp,Konoplya:2017wot,Bolokhov:2023dxq} . The correspondence is expected to be broken for gravitational perturbations of the Einstein-dilaton-Gauss-Bonnet theory \cite{Konoplya:2019hml}.  Here we will check whether this correspondence is fulfilled for the test fields in the EsGB theory.

\subsection{Quadratic GB-Coupling Functional: \texorpdfstring{$f(\varphi)=\varphi^2$}{TEXT}}\label{sec:3.1}

\begin{figure}[H]
\centering
\subfloat[$l=1$ \& $s=0$]{\includegraphics[width=.5\textwidth]{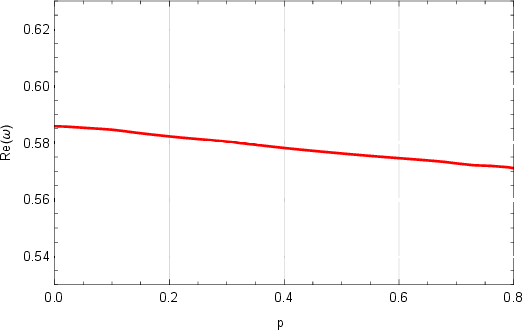}}\hfill
\subfloat[$l=1$ \& $s=0$]{\includegraphics[width=.5\textwidth]{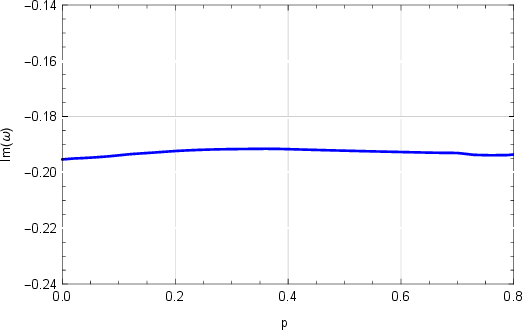}}\hfill
\caption{Fundamental quasinormal mode $(n=0)$ for massless scalar fields with 
$r_0=1$; red line denotes real part of frequency and blue lines denote the 
imaginary part of the frequency. }\label{fig:3}
\end{figure}

\begin{figure}[H]
\centering
\subfloat[$l=2$ \& $s=0$]{\includegraphics[width=.5\textwidth]{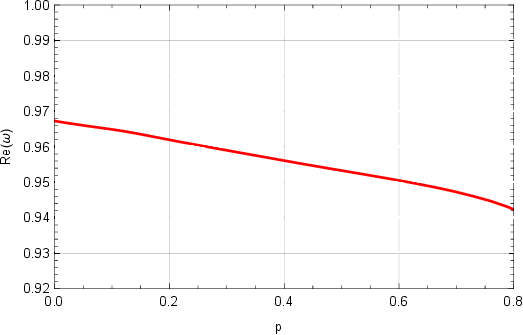}}\hfill
\subfloat[$l=2$ \& $s=0$]{\includegraphics[width=.5\textwidth]{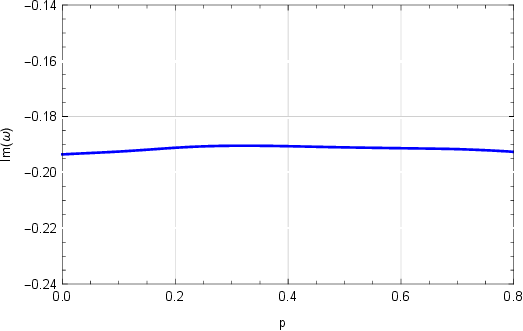}}\hfill
\caption{Fundamental quasinormal mode $(n=0)$ for massless scalar 
fields with $r_0=1$; red line denotes real part of frequency and 
blue lines denote the imaginary part of the frequency. }\label{fig:4}
\end{figure}

\begin{figure}[H]
\centering
\subfloat[$l=1$ \& $s=1$]{\includegraphics[width=.5\textwidth]{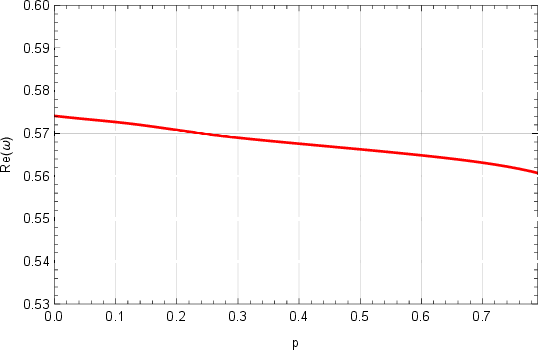}}\hfill
\subfloat[$l=1$ \& $s=1$]{\includegraphics[width=.5\textwidth]{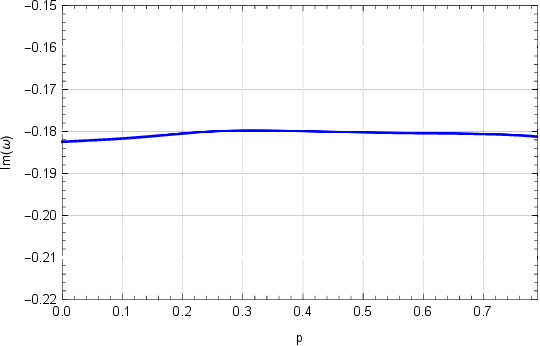}}\hfill
\caption{Fundamental quasinormal mode $(n=0)$ for electromagnetic 
fields with $r_0=1$; red line denotes real part of frequency and 
blue lines denote imaginary part of the frequency.}\label{fig:5}
\end{figure}

\begin{figure}[H]
\centering
\subfloat[$l=2$ \& $s=1$]{\includegraphics[width=.5\textwidth]{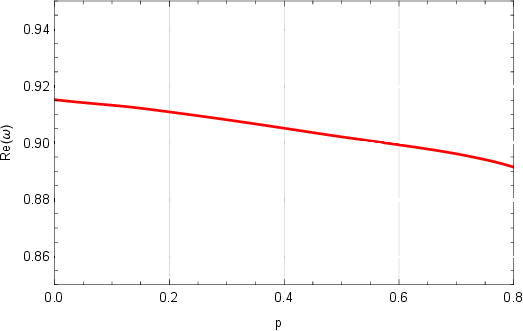}}\hfill
\subfloat[$l=2$ \& $s=1$]{\includegraphics[width=.5\textwidth]{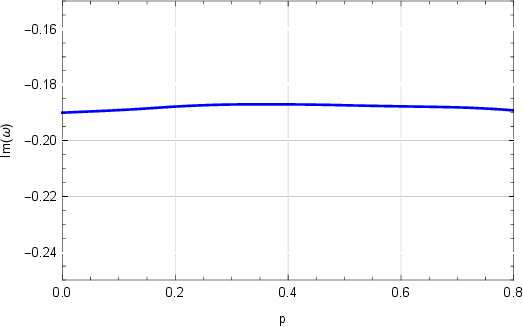}}\hfill
\caption{Fundamental quasinormal mode $(n=0)$ for electromagnetic 
fields with $r_0=1$; red line denotes real part of frequency and 
blue lines denote imaginary part of the frequency.}\label{fig:6}
\end{figure}

\begin{figure}[H]
\centering
\subfloat[$l=3$ \& $s=1$]{\includegraphics[width=.5\textwidth]{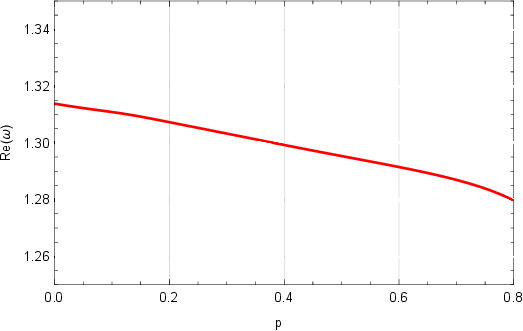}}\hfill
\subfloat[$l=3$ \& $s=1$]{\includegraphics[width=.5\textwidth]{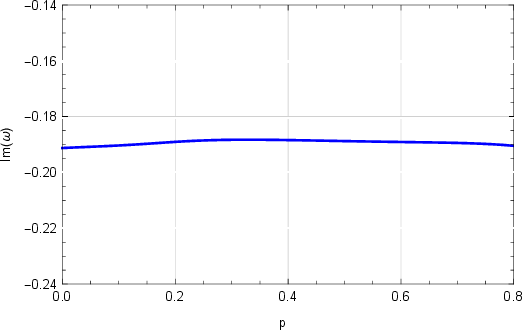}}\hfill
\caption{Fundamental quasinormal mode $(n=0)$ for electromagnetic 
fields with $r_0=1$; red line denotes real part of frequency and blue 
lines denote imaginary part of the frequency.}\label{fig:7}
\end{figure}

The quasinormal modes of massless scalar $(s=0)$ and electromagnetic $(s=1)$ 
fields for $l=1$, $2$ and $l=1$, $2$, $3$ are shown in Figs. \ref{fig:3}-\ref{fig:7}.
From Figs. \ref{fig:3}-\ref{fig:7} one can be seen that the behaviour of real and imaginary parts
of the quasinormal frequency as a function of parameter $p$ is almost 
linear for different values of $l$. Therefore, we use in-build 
\textit{Mathematica} function \textit{FindFormula}, to express real and 
imaginary parts of the quasinormal frequency as a function of parameter $p$
for different values of $l$. In formula \ref{eq:12} and \ref{eq:13}  
the parameter  $p$ runs from $p=0.0$ to $p=0.8$. The approximate linear laws of scalar and 
electromagnetic fields for different values of $l$ are given by

\begin{subequations}\label{eq:12} 
\begin{align}
    \text{Re}(\omega_{s=0, l=1}) &\approx 0.586146 -0.0192317 p, \label{eq:12a} \\
    \text{Im}(\omega_{s=0, l=1}) &\approx 0.0114263 p-0.194999, \\
    \text{Re}(\omega_{s=0, l=2}) &\approx 0.967948 -0.0299163 p, \label{eq:12c} \\
    \text{Im}(\omega_{s=0, l=2}) &\approx 0.0108174 p - 0.193500.
\end{align}
\end{subequations}

\begin{subequations}\label{eq:13} 
    \begin{align}
        \text{Re}(\omega_{s=1, l=1}) &\approx 0.498029 -0.0176951 p, \label{eq:13a} \\
        \text{Im}(\omega_{s=1, l=1}) &\approx 0.00835913 p-0.184913, \\
        \text{Re}(\omega_{s=1, l=2}) &\approx 0.916163 -0.0281436 p, \label{eq:13c} \\
        \text{Im}(\omega_{s=1, l=2}) &\approx 0.00940289 p-0.189931, \\
        \text{Re}(\omega_{s=1, l=3}) &\approx 1.31512 -0.0405207 p, \label{eq:13e} \\
        \text{Im}(\omega_{s=1, l=3}) &\approx 0.0103759 p-0.191264.
    \end{align}
    \end{subequations}

\begin{figure}[H]
    \centering
    \subfloat[$l=15$ \& $s=0$]{\includegraphics[width=.5\textwidth]{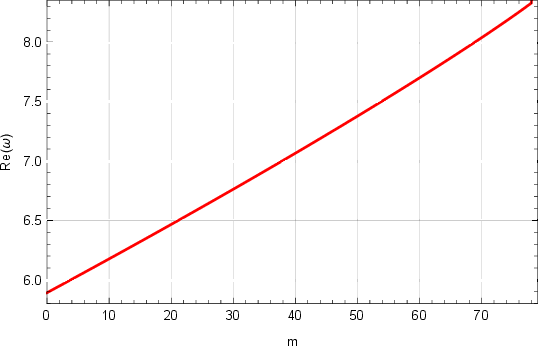}}\hfill
    \subfloat[$l=15$ \& $s=0$]{\includegraphics[width=.5\textwidth]{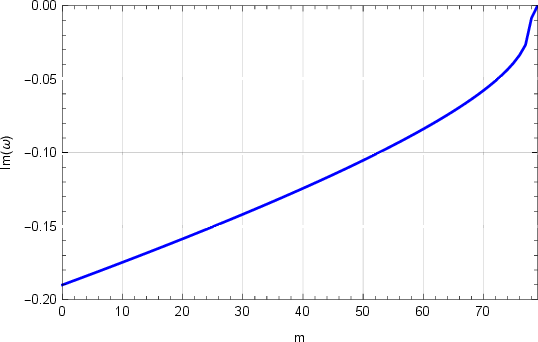}}\hfill
    \caption{The dependence of quasinormal frequency on mass $(m)$ for massive scalar fields with 
    $r_0=1$, $p=0.5$, red line denotes real part of frequency and blue lines denote imaginary 
    part of the frequency.}\label{fig:8}
\end{figure}

The behaviour of real and imaginary parts of the quasinormal modes for massive scalar 
fields as a function of mass ($m$) is shown in Fig. \ref{fig:8}. The WKB formula 
\cite{Schutz:1985km,Iyer:1986np,Konoplya:2003ii,Konoplya:2004ip} can not
describe the quasi resonances accurately \cite{Konoplya:2017tvu}, but in the eikonal 
regime WKB method is exact, providing sufficiently accurate results at $l \geq n$. Numerous examples of usage of the WKB  method (see, for instance \cite{Bolokhov:2023bwm,Konoplya:2019hlu,Zhao:2022gxl}) say that, 
the sixth-order WKB formula with Pad\'e approximants is usually the best for the computation of the quasinormal frequency of massive scalar fields. From Fig. \ref{fig:8}(b)
we can see that the imaginary part of the quasinormal frequency approaches 
zero, which indicates the existence of arbitrarily long-lived frequencies, called quasi-resonances.

\subsubsection{Analytical Formula For QNMS in the Eikonal Regime}

In the domain of high multipole numbers $l$ (eikonal), test fields with 
varying spin follows a common law up to the leading order. In this context, 
we examine an electromagnetic field with the effective potential 
\ref{eq:9}(b). In the eikonal regime, one can use the first-order WKB formula 
given by
\begin{equation}\label{eq:14} 
    \omega = \sqrt{V_{0} - i \Bigl( n+ \frac{1}{2}\Bigl) \sqrt{-2V_{0}^{\prime \prime}}},
\end{equation}
where $n=0$, $1$, $2$ is the overtone number, $V_{0}$ is the effective potential
at $r=r_{max}$ and $V_{0}^{\prime \prime}$ is the double derivative of effective potential
with respect to radial coordinate $r$, evaluated at $r=r_{max}$. Using
the Ref. \cite{Konoplya:2023moy} we find that the maximum of effective
potential occurs at $r=r_{max}$ which is given by  
\begin{equation}\label{eq:15} 
    r_{max}= \frac{3r_{0}}{2} +0.0387 r_{0} p + \mathcal{O}(p^2).
\end{equation}

Now, substituting $r_{max}$ and equation \ref{eq:9}(b) into equation \eqref{eq:14} one can obtain 
the analytical expression for quasinormal frequency in the eikonal regime as

\begin{equation}\label{eq:16} 
    \omega = \frac{(1+2l) (1-0.0258 p) - i (1+2n) (1-0.0447 p) }{3 \sqrt{3}r_{0}}.
\end{equation}
In the limit $p \to 0$, we obtained the analytical expression of quasinormal frequency
for Schwarzschild black hole. 

\subsection{Cubic GB-Coupling Functional: \texorpdfstring{$f(\varphi)=\varphi^3$}{TEXT}}\label{sec:3.2}

\begin{figure}[H]
\centering
\subfloat[$l=1$ \& $s=0$]{\includegraphics[width=.5\textwidth]{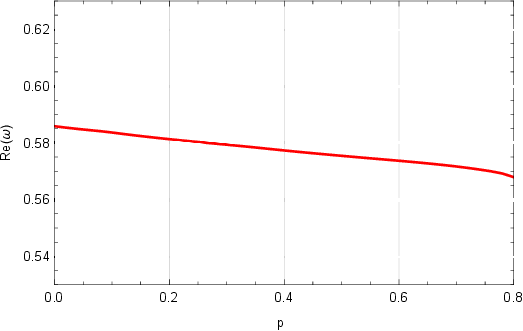}}\hfill
\subfloat[$l=1$ \& $s=0$]{\includegraphics[width=.5\textwidth]{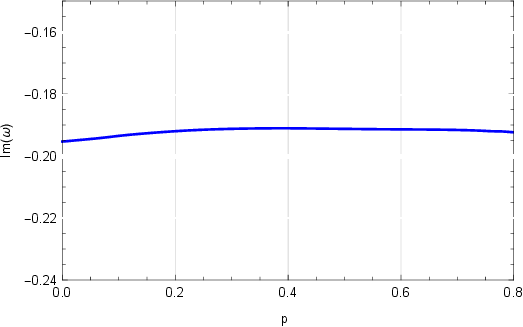}}\hfill
\caption{Fundamental quasinormal mode $(n=0)$ for massless scalar fields with 
$r_0=1$, red line denotes the real part of frequency and blue lines denote the 
imaginary part of the frequency. }\label{fig:9}
\end{figure}

\begin{figure}[H]
\centering
\subfloat[$l=2$ \& $s=0$]{\includegraphics[width=.5\textwidth]{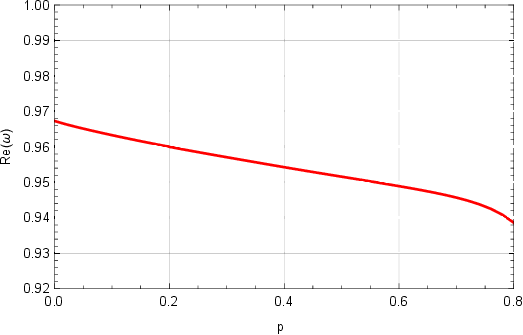}}\hfill
\subfloat[$l=2$ \& $s=0$]{\includegraphics[width=.5\textwidth]{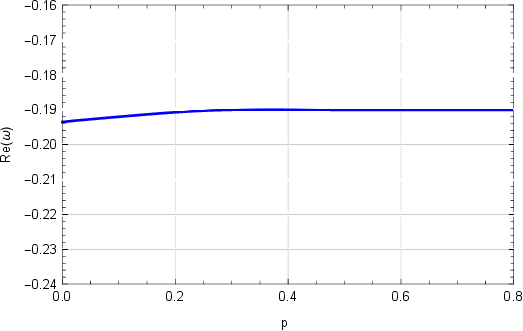}}\hfill
\caption{Fundamental quasinormal mode $(n=0)$ for massless scalar fields 
with $r_0=1$, red line denotes the real part of frequency and blue lines 
denote the imaginary part of the frequency. }\label{fig:10}
\end{figure}

\begin{figure}[H]
\centering
\subfloat[$l=1$ \& $s=1$]{\includegraphics[width=.5\textwidth]{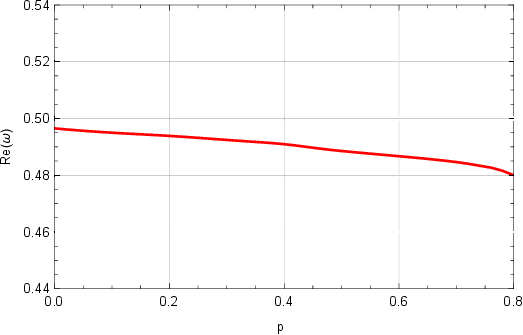}}\hfill
\subfloat[$l=1$ \& $s=1$]{\includegraphics[width=.5\textwidth]{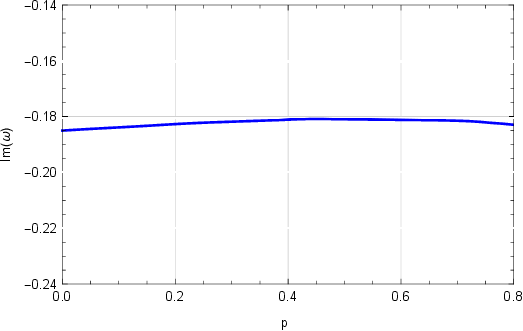}}\hfill
\caption{Fundamental quasinormal mode $(n=0)$ for electromagnetic 
fields with $r_0=1$, red line denotes real part of frequency and 
blue lines denote the imaginary part of the frequency.}\label{fig:11}
\end{figure}

\begin{figure}[H]
\centering
\subfloat[$l=2$ \& $s=1$]{\includegraphics[width=.5\textwidth]{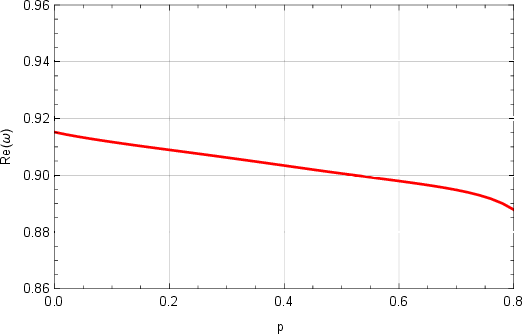}}\hfill
\subfloat[$l=2$ \& $s=1$]{\includegraphics[width=.5\textwidth]{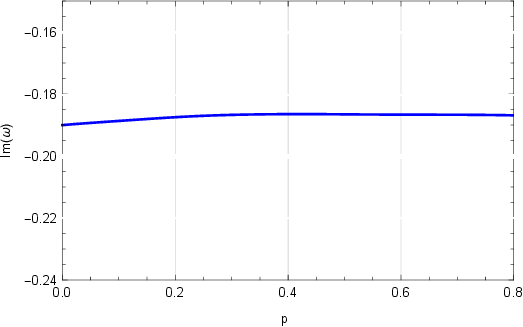}}\hfill
\caption{Fundamental quasinormal mode $(n=0)$ for electromagnetic 
fields with $r_0=1$, red line denotes real part of frequency and 
blue lines denote the imaginary part of the frequency.}\label{fig:12}
\end{figure}

\begin{figure}[H]
\centering
\subfloat[$l=3$ \& $s=1$]{\includegraphics[width=.5\textwidth]{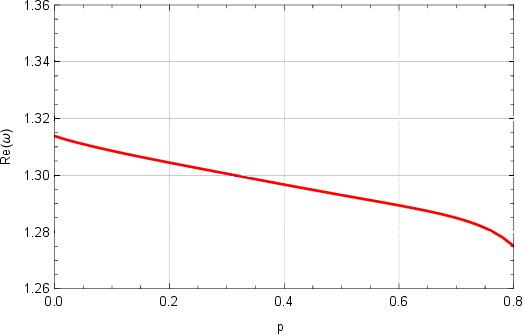}}\hfill
\subfloat[$l=3$ \& $s=1$]{\includegraphics[width=.5\textwidth]{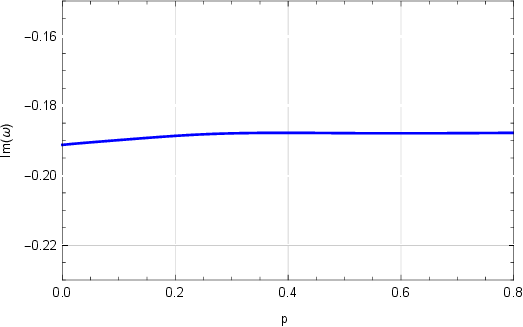}}\hfill
\caption{Fundamental quasinormal mode $(n=0)$ for electromagnetic 
fields with $r_0=1$, red line denotes real part of frequency and 
blue lines denote the imaginary part of the frequency.}\label{fig:13}
\end{figure}

The quasinormal modes of massless scalar $(s=0)$ and electromagnetic $(s=1)$ 
fields for $l=1$, $2$ and $l=1$, $2$, $3$ are shown in Figs. \ref{fig:9}-\ref{fig:13}.
From Figs. \ref{fig:9}-\ref{fig:13} one can be seen that the behaviour of real and imaginary parts
of the quasinormal frequency as a function of parameter $p$ is almost 
linear for different values of $l$. In formula \ref{eq:17} and \ref{eq:18} 
the parameter $p$ runs from $p=0.0$ to $p=0.6$. After $p=0.6$, the behaviour of real and imaginary parts of quasinormal frequency is non-linear, especially for $l=2$ (scalar fields, Fig. \ref{fig:10}) and $l=2$, $3$ (electromagnetic fields, Figs. \ref{fig:12} \& \ref{fig:13}). The approximated linear laws of scalar and 
electromagnetic fields for different values of $l$ are given by

\begin{subequations}\label{eq:17} 
\begin{align}
    \text{Re}(\omega_{s=0, l=1}) &\approx 0.586345 -0.0231319 p, \label{eq:17a} \\
    \text{Im}(\omega_{s=0, l=1}) &\approx 0.0236755 p-0.258812, \\
    \text{Re}(\omega_{s=0, l=2}) &\approx 0.967896 -0.0355368 p, \label{eq:17c} \\
    \text{Im}(\omega_{s=0, l=2}) &\approx 0.0114912 p-0.193244.
\end{align}
\end{subequations}

\begin{subequations}\label{eq:18} 
    \begin{align}
        \text{Re}(\omega_{s=1, l=1}) &\approx 0.498397 -0.0219789 p, \label{eq:18a} \\
        \text{Im}(\omega_{s=1, l=1}) &\approx 0.0134254p - 0.202685, \\
        \text{Re}(\omega_{s=1, l=2}) &\approx 0.916381 - 0.0344771 p, \label{eq:18c} \\
        \text{Im}(\omega_{s=1, l=2}) &\approx 0.0147288 p - 0.239645, \\
        \text{Re}(\omega_{s=1, l=3}) &\approx 1.31503 - 0.0482737 p, \label{eq:18e} \\
        \text{Im}(\omega_{s=1, l=3}) &\approx 0.0174108 p - 0.233444.
    \end{align}
\end{subequations}

\begin{figure}[H]
    \centering
    \subfloat[$l=15$ \& $s=0$]{\includegraphics[width=.5\textwidth]{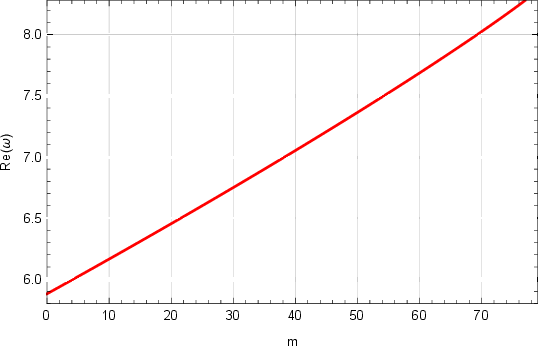}}\hfill
    \subfloat[$l=15$ \& $s=0$]{\includegraphics[width=.5\textwidth]{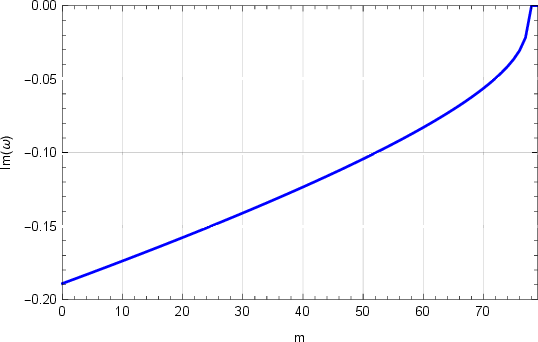}}\hfill
    \caption{The dependence of quasinormal frequency on mass $(m)$ for massive 
    scalar fields with $r_0=1$, $p=0.5$ red line denote the real part of frequency and 
    blue lines denote the imaginary part of the frequency. }\label{fig:14}
\end{figure}

The behaviour of real and imaginary parts of the quasinormal modes for massive scalar 
fields as a function of mass ($m$) is shown in Fig. \ref{fig:14}. From Fig. \ref{fig:14}(a) we can see that the real part of quasinormal frequency is an increasing function of mass $m$ and the imaginary part of the quasinormal frequency (Fig. \ref{fig:14}b) approaches zero, which indicates the existence of quasi-resonances. 

\subsubsection{Analytical Formula For QNMS in the Eikonal Regime}

In the domain of high multipole numbers $l$ (eikonal), test fields with 
varying spin follow a common law up to the leading order. In this context, 
we examine an electromagnetic field with the effective potential 
\ref{eq:9}(b). In the eikonal regime, one can use the first-order WKB formula 
given in equation \eqref{eq:14}. With the help of Ref. \cite{Konoplya:2023moy} we find that the maximum of effective
potential occurs at $r=r_{max}$ which is given by  
\begin{equation}\label{eq:20}
    r_{max}= \frac{3r_{0}}{2} +0.0637 r_{0} p + \mathcal{O}(p^2).
\end{equation}
Now, substituting $r_{max}$ and equation \ref{eq:9}(b) into equation \eqref{eq:14} one can obtain the analytical expression for quasinormal frequency in the eikonal regime for cubic GB-coupling functional as
\begin{equation}\label{eq:21}
    \omega = \frac{(1+2l) (1-0.0596 p) - i (1+2n) (1-0.1071 p) }{3 \sqrt{3}r_{0}}.
\end{equation}
In the limit $p \to 0$, we obtained the analytical expression of quasinormal frequency
for Schwarzschild black hole.

\subsection{Quartic GB-Coupling Functional: \texorpdfstring{$f(\varphi)=\varphi^4$}{TEXT}}\label{sec:3.3}

\begin{figure}[H]
\centering
\subfloat[$l=1$ \& $s=0$]{\includegraphics[width=.5\textwidth]{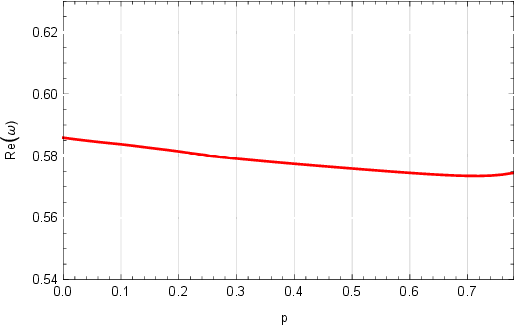}}\hfill
\subfloat[$l=1$ \& $s=0$]{\includegraphics[width=.5\textwidth]{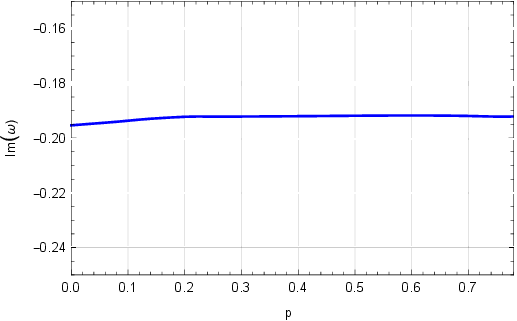}}\hfill
\caption{Fundamental quasinormal mode $(n=0)$ for massless scalar fields 
with $r_0=1$, red line denotes the real part of frequency and blue lines 
denote the imaginary part of the frequency. }\label{fig:15}
\end{figure}

\begin{figure}[H]
\centering
\subfloat[$l=2$ \& $s=0$]{\includegraphics[width=.5\textwidth]{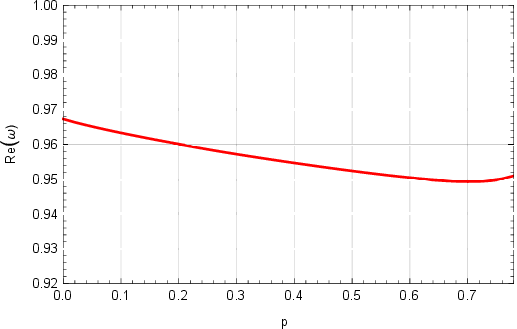}}\hfill
\subfloat[$l=2$ \& $s=0$]{\includegraphics[width=.5\textwidth]{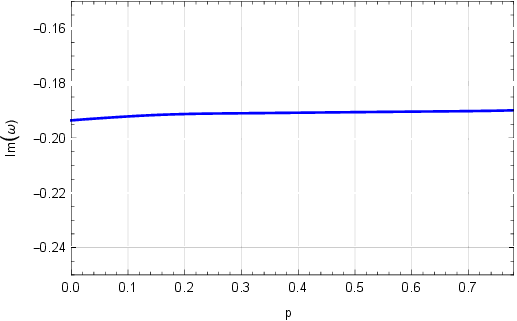}}\hfill
\caption{Fundamental quasinormal mode $(n=0)$ for 
 massless scalar fields with $r_0=1$, red line denotes the real part 
 of frequency and blue lines denote the imaginary part of the frequency. }\label{fig:16}
\end{figure}

\begin{figure}[H]
\centering
\subfloat[$l=1$ \& $s=1$]{\includegraphics[width=.5\textwidth]{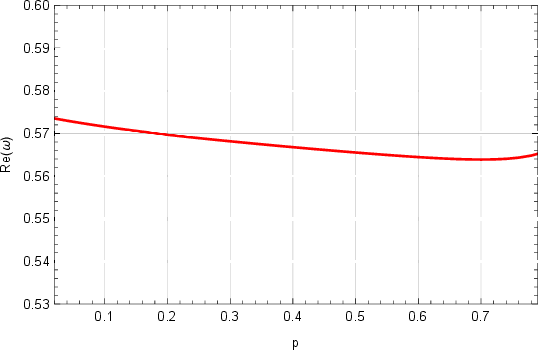}}\hfill
\subfloat[$l=1$ \& $s=1$]{\includegraphics[width=.5\textwidth]{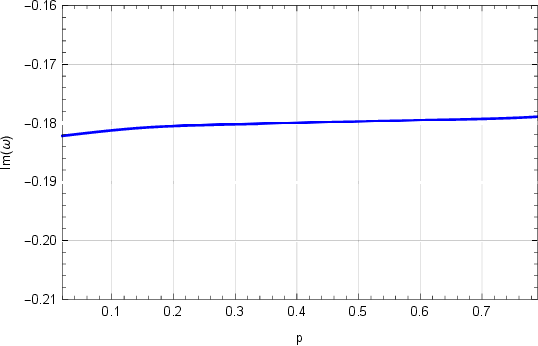}}\hfill
\caption{Fundamental quasinormal mode $(n=0)$ for electromagnetic fields 
with $r_0=1$, red line denotes real part of frequency and blue lines denote 
the imaginary part of the frequency.}\label{fig:17}
\end{figure}

\begin{figure}[H]
\centering
\subfloat[$l=2$ \& $s=1$]{\includegraphics[width=.5\textwidth]{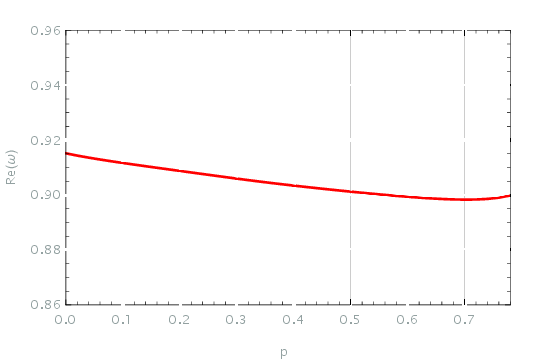}}\hfill
\subfloat[$l=2$ \& $s=1$]{\includegraphics[width=.5\textwidth]{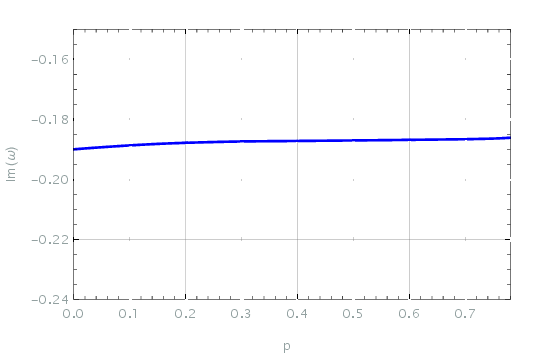}}\hfill
\caption{Fundamental quasinormal mode $(n=0)$ for electromagnetic 
fields with $r_0=1$, red line denotes real part of frequency and blue 
lines denote the imaginary part of the frequency.}\label{fig:18}
\end{figure}

\begin{figure}[H]
\centering
\subfloat[$l=3$ \& $s=1$]{\includegraphics[width=.5\textwidth]{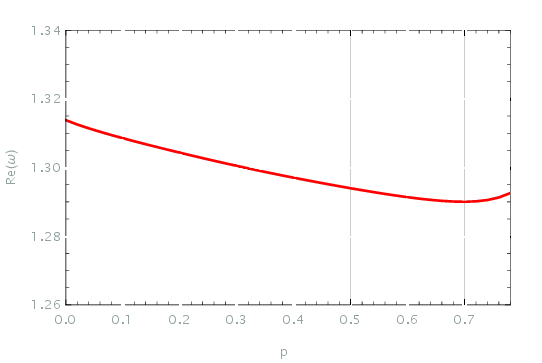}}\hfill
\subfloat[$l=3$ \& $s=1$]{\includegraphics[width=.5\textwidth]{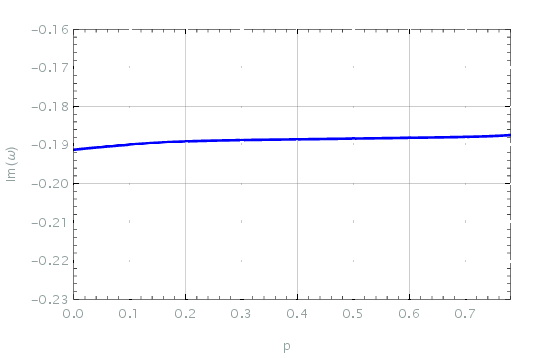}}\hfill
\caption{Fundamental quasinormal mode $(n=0)$ for electromagnetic fields 
with $r_0=1$, red line denotes real part of frequency and blue lines denote 
the imaginary part of the frequency.}\label{fig:19}
\end{figure}

The quasinormal modes of massless scalar $(s=0)$ and electromagnetic $(s=1)$ 
fields for $l=1$, $2$ and $l=1$, $2$, $3$ are shown in Figs. \ref{fig:15}-\ref{fig:19}.
From Figs. \ref{fig:15}-\ref{fig:19} one can be seen that the behaviour of real and imaginary parts
of the quasinormal frequency as a function of parameter $p$ is almost 
linear for different values of $l$. In formula \ref{eq:22} and \ref{eq:23}
the approximated analytical expression for quasinormal frequency is written down for different values of $l$, where parameter $p$ runs from $p=0.0$ to $p=0.7$. After $p=0.7$, the non-linear behaviour is visible. The approximated linear laws of scalar and 
electromagnetic fields for different values of $l$ are given by

\begin{subequations}\label{eq:22} 
\begin{align}
    \text{Re}(\omega_{s=0, l=1}) &\approx 0.583828 -0.0130698 p, \label{eq:22a} \\
    \text{Im}(\omega_{s=0, l=1}) &\approx 0.0230895 p-0.272239, \\
    \text{Re}(\omega_{s=0, l=2}) &\approx 1.00388 -0.0383934 p, \label{eq:22c} \\
    \text{Im}(\omega_{s=0, l=2}) &\approx 0.00397625 p-0.192472.
\end{align}
\end{subequations}

\begin{subequations}\label{eq:23} 
    \begin{align}
        \text{Re}(\omega_{s=1, l=1}) &\approx 0.495561 -0.0117364 p, \label{eq:23a} \\
        \text{Im}(\omega_{s=1, l=1}) &\approx 0.00632363 p-0.184499, \\
        \text{Re}(\omega_{s=1, l=2}) &\approx 0.943469 -0.0345489 p, \label{eq:23c} \\
        \text{Im}(\omega_{s=1, l=2}) &\approx 0.00445784 p-0.189155, \\
        \text{Re}(\omega_{s=1, l=3}) &\approx 1.36289 -0.0510792 p, \label{eq:23e} \\
        \text{Im}(\omega_{s=1, l=3}) &\approx 0.00418319 p-0.190336.
    \end{align}
\end{subequations}

\begin{figure}[H]
    \centering
    \subfloat[$l=15$ \& $s=0$]{\includegraphics[width=.5\textwidth]{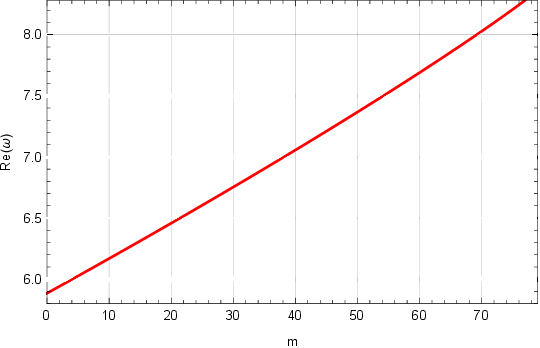}}\hfill
    \subfloat[$l=15$ \& $s=0$]{\includegraphics[width=.5\textwidth]{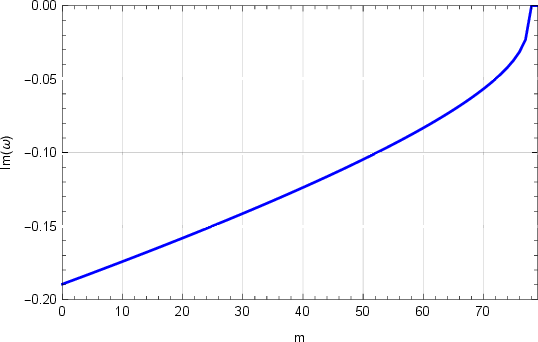}}\hfill
    \caption{The dependence of quasinormal frequency on mass $(m)$ for massive fields with 
    $r_0=1$, $p=0.5$ red line denotes real part of frequency and blue lines denote the
     imaginary part of the frequency.}\label{fig:20}
\end{figure}

The behaviour of real and imaginary parts of the quasinormal modes for massive scalar fields as a function of mass ($m$) is shown in Fig. \ref{fig:20} for quartic GB-coupling functional. The real part of the quasinormal frequency is shown in Fig. \ref{fig:20}(a), which is an increasing function of mass $m$. The imaginary part is shown in Fig. \ref{fig:20}(b) and it approaches zero as mass increases.
Therefore, the behaviour of the imaginary part of quasinormal frequency indicates the existence of quasi-resonances for quartic GB-coupling functional.

\subsubsection{Analytical Formula For QNMS in the Eikonal Regime}

In the domain of high multipole numbers $l$ (eikonal), test fields with 
varying spin follow a common law up to the leading order. In this context, 
we examine an electromagnetic field with the effective potential 
\ref{eq:9}(b). In the eikonal regime, one can use the first-order WKB formula 
given in equation \eqref{eq:14}. With the help of numerical analysis given in Ref. \cite{Konoplya:2023moy} we find that the maximum of effective potential occurs at $r=r_{max}$ which is given by 

\begin{equation}\label{eq:25}
    r_{max}= \frac{3r_{0}}{2} +0.0527 r_{0} p + \mathcal{O}(p^2).
\end{equation}

Now, substituting $r_{max}$ and equation \ref{eq:9}(b) into equation \eqref{eq:14} one can obtain 
the analytical expression for quasinormal frequency in the eikonal regime as

\begin{equation}\label{eq:26}
    \omega = \frac{(1+2l) (1-0.0570 p) - i (1+2n) (1-0.0971 p) }{3 \sqrt{3}r_{0}}.
\end{equation}

In the limit $p \to 0$, we obtained the analytical expression of quasinormal frequency
for Schwarzschild black hole.

\subsection{ Inverse GB-Coupling Functional: \texorpdfstring{$f(\varphi)=\varphi{-1}$}{TEXT}}\label{sec:3.4}

\begin{figure}[H]
\centering
\subfloat[$l=1$ \& $s=0$]{\includegraphics[width=.5\textwidth]{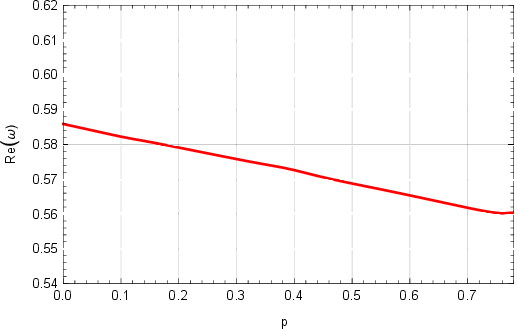}}\hfill
\subfloat[$l=1$ \& $s=0$]{\includegraphics[width=.5\textwidth]{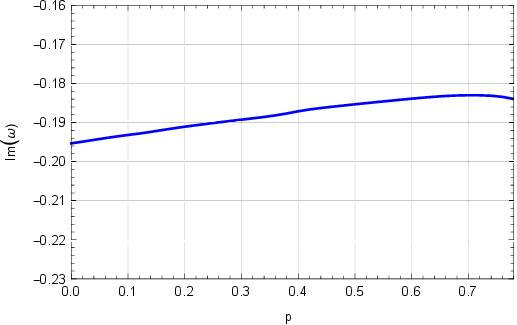}}\hfill
\caption{Fundamental quasinormal mode $(n=0)$ for massless scalar fields 
with $r_0=1$, red line denotes the real part of frequency and blue lines 
denote the imaginary part of the frequency. }\label{fig:21}
\end{figure}

\begin{figure}[H]
\centering
\subfloat[$l=2$ \& $s=0$]{\includegraphics[width=.5\textwidth]{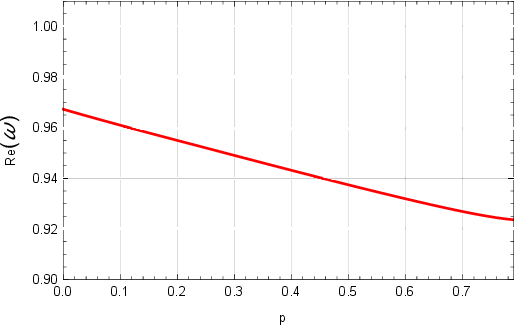}}\hfill
\subfloat[$l=2$ \& $s=0$]{\includegraphics[width=.5\textwidth]{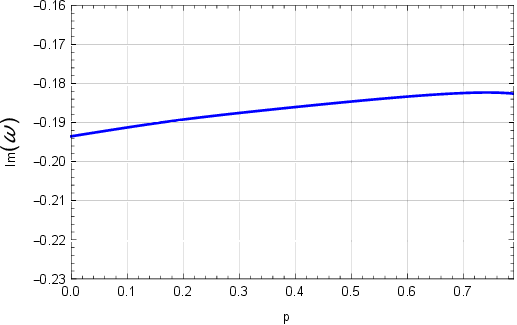}}\hfill
\caption{Fundamental quasinormal mode $(n=0)$ for massless scalar fields 
with $r_0=1$, red line denotes the real part of frequency and blue lines 
denote the imaginary part of the frequency. }\label{fig:22}
\end{figure}

\begin{figure}[H]
\centering
\subfloat[$l=1$ \& $s=1$]{\includegraphics[width=.5\textwidth]{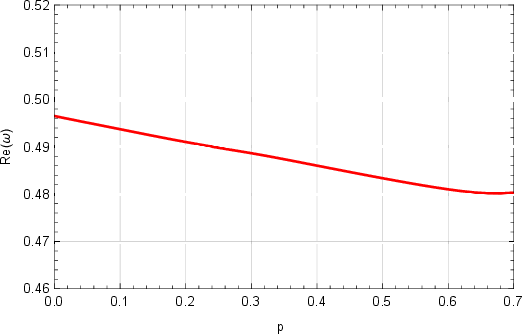}}\hfill
\subfloat[$l=1$ \& $s=1$]{\includegraphics[width=.5\textwidth]{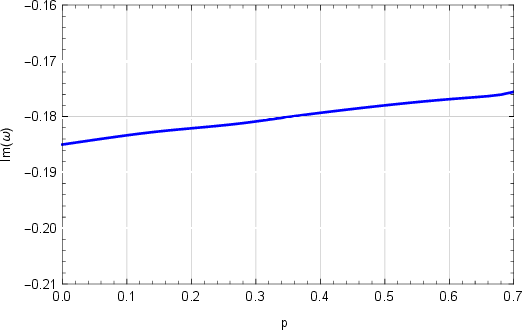}}\hfill
\caption{Fundamental quasinormal mode $(n=0)$ for electromagnetic 
fields with $r_0=1$, red line denotes real part of frequency and blue 
lines denote the imaginary part of the frequency.}\label{fig:23}
\end{figure}

\begin{figure}[H]
\centering
\subfloat[$l=2$ \& $s=1$]{\includegraphics[width=.5\textwidth]{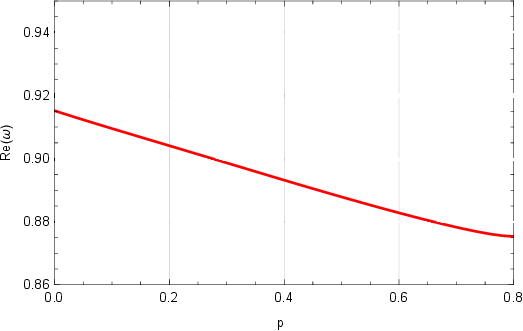}}\hfill
\subfloat[$l=2$ \& $s=1$]{\includegraphics[width=.5\textwidth]{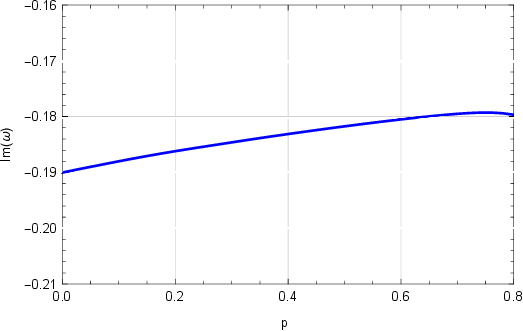}}\hfill
\caption{Fundamental quasinormal mode $(n=0)$ for electromagnetic 
fields with $r_0=1$, red line denotes real part of frequency and blue 
lines denote the imaginary part of the frequency.}\label{fig:24}
\end{figure}

\begin{figure}[H]
\centering
\subfloat[$l=3$ \& $s=1$]{\includegraphics[width=.5\textwidth]{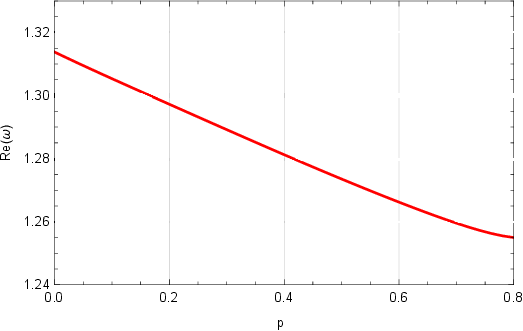}}\hfill
\subfloat[$l=3$ \& $s=1$]{\includegraphics[width=.5\textwidth]{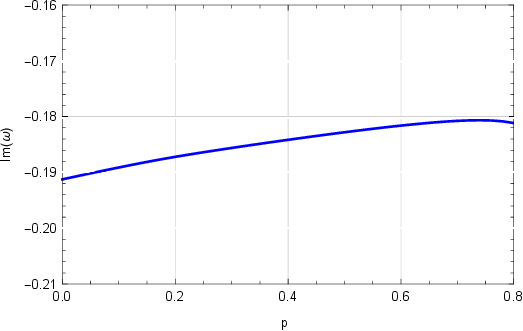}}\hfill
\caption{Fundamental quasinormal mode $(n=0)$ for electromagnetic 
fields with $r_0=1$, red line denotes real part of frequency and blue 
lines denote the imaginary part of the frequency.}\label{fig:25}
\end{figure}

The quasinormal modes of massless scalar $(s=0)$ and electromagnetic $(s=1)$ 
fields for $l=1$, $2$ and $l=1$, $2$, $3$ are shown in Figs. \ref{fig:21}-\ref{fig:25}.
From Figs. \ref{fig:21}-\ref{fig:25} one can see that the behaviour of real and imaginary parts
of the quasinormal frequency as a function of parameter $p$ is almost 
linear for different values of $l$. In formula \ref{eq:27} and \ref{eq:28} 
the parameter $p$ runs from $p=0.0$ to $p=0.65$. The approximated linear law of scalar and 
electromagnetic fields for different values of $l$ are given by

\begin{subequations}\label{eq:27} 
    \begin{align}
        \text{Re}(\omega_{s=0, l=1}) &\approx 0.58558 -0.0330224 p, \label{eq:27a} \\
        \text{Im}(\omega_{s=0, l=1}) &\approx 0.0183859 p-0.194868, \\
        \text{Re}(\omega_{s=0, l=2}) &\approx 0.966147 -0.0558721 p, \label{eq:27c} \\
        \text{Im}(\omega_{s=0, l=2}) &\approx 0.0231498 p-0.215501.
    \end{align}
    \end{subequations}

    \begin{subequations}\label{eq:28} 
        \begin{align}
            \text{Re}(\omega_{s=1, l=1}) &\approx 0.495763 -0.0233709 p, \label{eq:28a} \\
            \text{Im}(\omega_{s=1, l=1}) &\approx 0.0139608 p-0.184918, \\
            \text{Re}(\omega_{s=1, l=2}) &\approx 0.915373 -0.0367994 p, \label{eq:28c} \\
            \text{Im}(\omega_{s=1, l=2}) &\approx 0.0122145 p - 0.188642, \\
            \text{Re}(\omega_{s=1, l=3}) &\approx 1.31159 -0.0728545 p, \label{eq:28e} \\
            \text{Im}(\omega_{s=1, l=3}) &\approx 0.011762 p-0.189603.
        \end{align}
    \end{subequations}

\begin{figure}[H]
    \centering
    \subfloat[$l=15$ \& $s=0$]{\includegraphics[width=.5\textwidth]{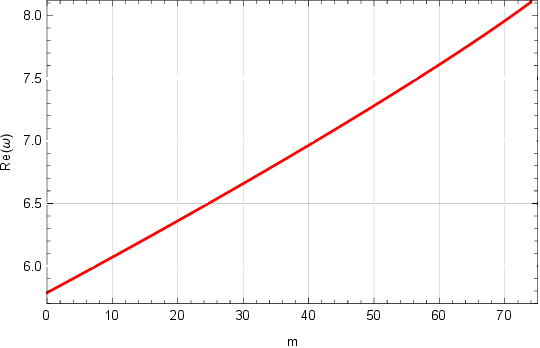}}\hfill
    \subfloat[$l=15$ \& $s=0$]{\includegraphics[width=.5\textwidth]{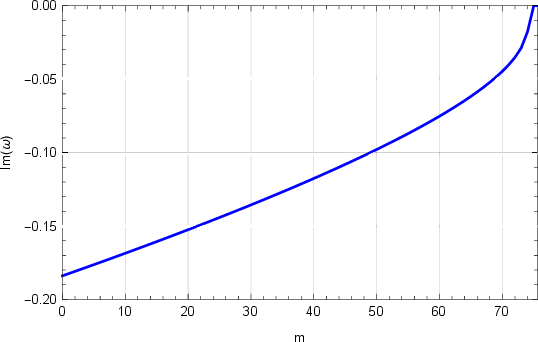}}\hfill
    \caption{The dependence of quasinormal frequency on mass $(m)$ for massive scalar 
    fields with $r_0=1$, $p=0.5$, red line denotes real part of frequency and blue 
    lines denote the imaginary part of the frequency.}\label{fig:26}
\end{figure}

The behaviour of real and imaginary parts of the quasinormal modes for massive scalar fields as a function of mass ($m$) is shown in Fig. \ref{fig:26} for inverse GB-coupling functional. From Fig. \ref{fig:26}(b)
we can see that the imaginary part of the quasinormal frequency slowly approaches 
zero as mass $m$ increases, which indicates the existence of arbitrarily long-lived frequencies, called quasi-resonances.

\subsubsection{Analytical Formula For QNMS in the Eikonal Regime}
In the domain of high multipole numbers $l$ (eikonal), test fields with 
varying spin follow a common law up to the leading order. In this context, 
we examine an electromagnetic field with the effective potential 
\ref{eq:9}(b). In the eikonal regime, one can use the first-order WKB formula 
given in equation \eqref{eq:14}. Using the Ref. \cite{Konoplya:2023moy} we find that the maximum of effective potential occurs at $r=r_{max}$ which is given by 

\begin{equation}
    r_{max}= \frac{3r_{0}}{2} +0.0563 r_{0} p + \mathcal{O}(p^2).
\end{equation}

Now, substituting $r_{max}$ and equation \ref{eq:9}(b) into equation \eqref{eq:14} one can obtain the analytical expression for quasinormal frequency in the eikonal regime as

\begin{equation}
    \omega = \frac{(1+2l) (1-0.0699 p) - i (1+2n) (1-0.1201 p) }{3 \sqrt{3}r_{0}}.
\end{equation}

In the limit $p \to 0$, we obtained the analytical expression of quasinormal frequency
for Schwarzschild black hole.

\subsection{Logarithmic GB-Coupling Functional: \texorpdfstring{$f(\varphi)=\ln{(\varphi)}$}{TEXT}}\label{sec:3.5}

\begin{figure}[H]
\centering
\subfloat[$l=1$ \& $s=0$]{\includegraphics[width=.5\textwidth]{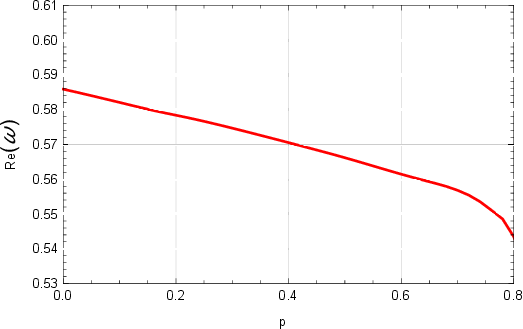}}\hfill
\subfloat[$l=1$ \& $s=0$]{\includegraphics[width=.5\textwidth]{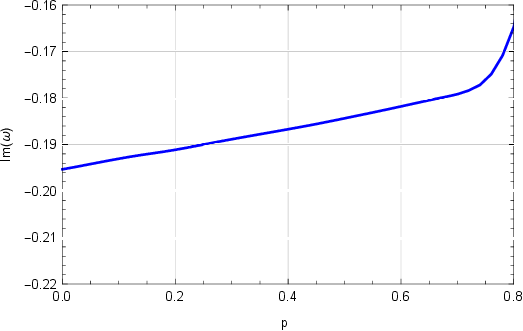}}\hfill
\caption{Fundamental quasinormal mode $(n=0)$ for massless scalar fields with 
$r_0=1$, red line denotes the real part of frequency and blue lines denote the 
imaginary part of the frequency. }\label{fig:27}
\end{figure}

\begin{figure}[H]
\centering
\subfloat[$l=2$ \& $s=0$]{\includegraphics[width=.5\textwidth]{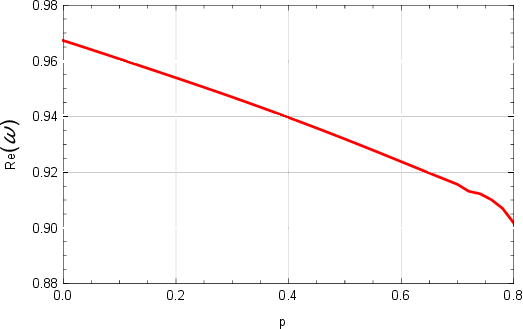}}\hfill
\subfloat[$l=2$ \& $s=0$]{\includegraphics[width=.5\textwidth]{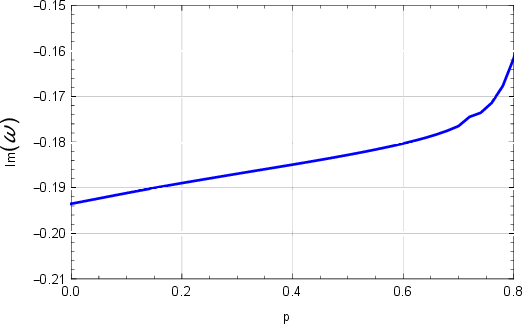}}\hfill
\caption{Fundamental quasinormal mode $(n=0)$ for massless scalar 
fields with $r_0=1$, red line denotes the real part of frequency and 
blue lines denote the imaginary part of the frequency. }\label{fig:28}
\end{figure}

\begin{figure}[H]
\centering
\subfloat[$l=1$ \& $s=1$]{\includegraphics[width=.5\textwidth]{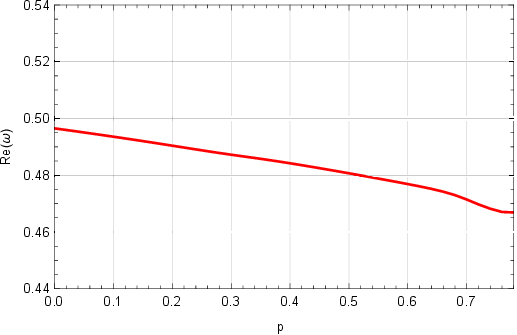}}\hfill
\subfloat[$l=1$ \& $s=1$]{\includegraphics[width=.5\textwidth]{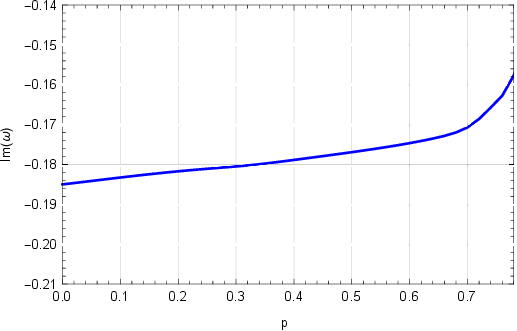}}\hfill
\caption{Fundamental quasinormal mode $(n=0)$ for electromagnetic 
fields with $r_0=1$, red line denotes real part of frequency and 
blue lines denote the imaginary part of the frequency.}\label{fig:29}
\end{figure}

\begin{figure}[H]
\centering
\subfloat[$l=2$ \& $s=1$]{\includegraphics[width=.5\textwidth]{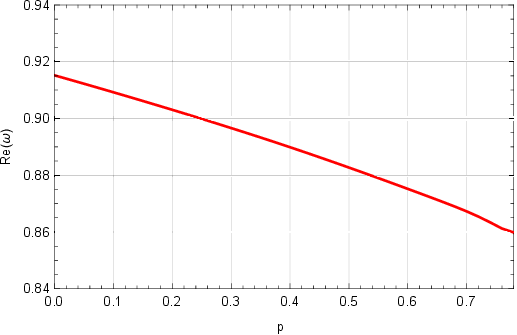}}\hfill
\subfloat[$l=2$ \& $s=1$]{\includegraphics[width=.5\textwidth]{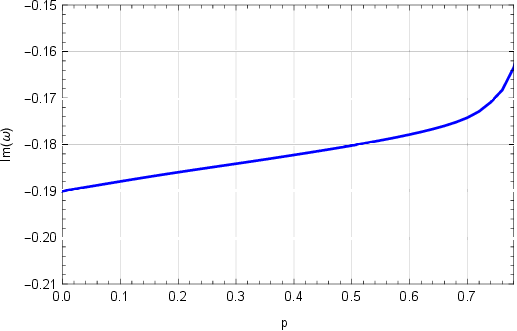}}\hfill
\caption{Fundamental quasinormal mode $(n=0)$ for electromagnetic 
fields with $r_0=1$, red line denotes real part of frequency and 
blue lines denote the imaginary part of the frequency.}\label{fig:30}
\end{figure}

\begin{figure}[H]
\centering
\subfloat[$l=3$ \& $s=1$]{\includegraphics[width=.5\textwidth]{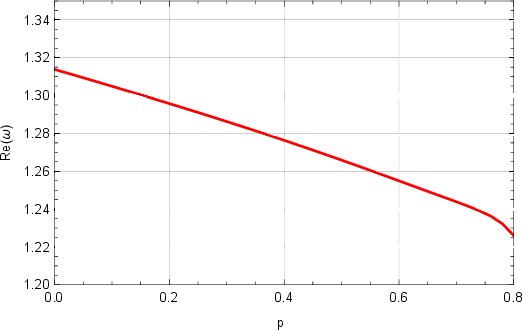}}\hfill
\subfloat[$l=3$ \& $s=1$]{\includegraphics[width=.5\textwidth]{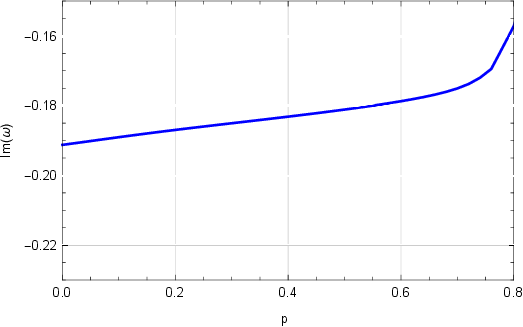}}\hfill
\caption{Fundamental quasinormal mode $(n=0)$ for electromagnetic fields 
with $r_0=1$, red line denotes real part of frequency and blue lines denote 
the imaginary part of the frequency.}\label{fig:31}
\end{figure}

The quasinormal modes of massless scalar $(s=0)$ and electromagnetic $(s=1)$ 
fields for $l=1$, $2$ and $l=1$, $2$, $3$ are shown in Figs. \ref{fig:27}-\ref{fig:31}.
From Figs. \ref{fig:27}-\ref{fig:31} one can see that the behaviour of real and imaginary parts
of the quasinormal frequency as a function of parameter $p$ is almost 
linear for different values of $l$. In formulas \ref{eq:29} and \ref{eq:30} 
the parameter $p$ runs from $p=0.0$ to $p=0.70$. The approximated linear laws for scalar and 
electromagnetic fields for different values of $l$ are given by

\begin{subequations}\label{eq:29} 
    \begin{align}
        \text{Re}(\omega_{s=0, l=1}) &\approx 0.587079 -0.043569 p, \label{eq:29a} \\
        \text{Im}(\omega_{s=0, l=1}) &\approx 0.0231869 p-0.195646, \\
        \text{Re}(\omega_{s=0, l=2}) &\approx 0.969141 -0.0766068 p, \label{eq:29c} \\
        \text{Im}(\omega_{s=0, l=2}) &\approx 0.0245713 p-0.19412.
    \end{align}
\end{subequations}

\begin{subequations}\label{eq:30} 
        \begin{align}
            \text{Re}(\omega_{s=1, l=1}) &\approx 0.497885 -0.0368047 p, \label{eq:33a} \\
            \text{Im}(\omega_{s=1, l=1}) &\approx 0.0271431 p-0.187726, \\
            \text{Re}(\omega_{s=1, l=2}) &\approx 0.915132 -0.0449956 p, \label{eq:33c} \\
            \text{Im}(\omega_{s=1, l=2}) &\approx 0.0277317 p-0.191956, \\
            \text{Re}(\omega_{s=1, l=3}) &\approx 1.31625 -0.103508 p, \label{eq:33e} \\
            \text{Im}(\omega_{s=1, l=3}) &\approx 0.0252856 p-0.192324.
        \end{align}
\end{subequations}

\begin{figure}[H]
    \centering
    \subfloat[$l=15$ \& $s=0$]{\includegraphics[width=.5\textwidth]{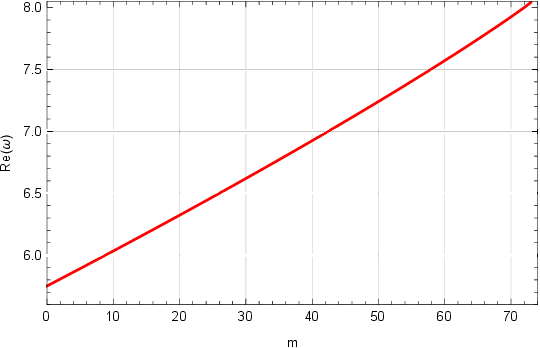}}\hfill
    \subfloat[$l=15$ \& $s=0$]{\includegraphics[width=.5\textwidth]{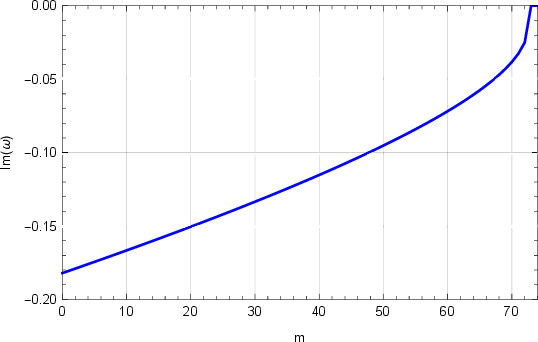}}\hfill
    \caption{The dependence of quasinormal frequency on mass $(m)$ for massive scalar fields with $r_0=1$, $p=0.5$, 
    red line denotes real part of frequency and blue lines denote the imaginary part 
    of the frequency.}\label{fig:32}
    \end{figure}

The behaviour of real and imaginary parts of the quasinormal modes for massive scalar fields as a function of mass ($m$) is shown in Fig. \ref{fig:32}. From Fig. \ref{fig:32}(b) we can see that the imaginary part of the quasinormal frequency approaches zero, which indicates the existence of quasi-resonances.

\subsubsection{Analytical Formula For QNMS in the Eikonal Regime}
In the domain of high multipole numbers $l$ (eikonal), test fields with 
varying spin follow a common law up to the leading order. In this context, 
we examine an electromagnetic field with the effective potential 
\ref{eq:9}(b). In the eikonal regime, one can use the first-order WKB formula 
given in equation \eqref{eq:14}. Using the Ref. \cite{Konoplya:2023moy} we find that the maximum of effective potential \ref{eq:9}(b) occurs at $r=r_{max}$ which is given by 

\begin{equation}\label{eq:35}
    r_{max}= \frac{3r_{0}}{2} +0.0544 r_{0} p + \mathcal{O}(p^2).
\end{equation}

Now, substituting $r_{max}$ and equation \ref{eq:9}(b) into equation \eqref{eq:14} one can obtain the analytical expression for quasinormal frequency in the eikonal regime for logarithmic GB-coupling functional as

\begin{equation}\label{eq:36}
    \omega = \frac{(1+2l) (1-0.0690 p) - i (1+2n) (1-0.1181 p) }{3 \sqrt{3}r_{0}}.
\end{equation}

In the limit $p \to 0$, we obtained the analytical expression of quasinormal frequency for Schwarzschild black hole. 

As one can see from this and previous analytical relations for the eikonal regime, the quasinormal modes are fully reproduced by the WKB formula and, therefore, the correspondence between eikonal qnms and null geodesics suggested in \cite{Cardoso:2008bp} is indeed valid for the test fields under consideration.

\section{Grey--Body Factors}\label{sec:4}

The strength of Hawking radiation does not fully reach a 
distant observer, it is partially suppressed by the effective 
potential surrounding the black holes and reflected to the 
black hole event horizon. To find the number of particles 
reflected due to the effective potential, it is essential to 
determine the grey--body factors. This involves solving the classical 
scattering problem to estimate the number of particles that undergo 
reflection.

We will investigate the wave equation \eqref{eq:7} under the 
boundary conditions that allow the inclusion of incoming 
waves from infinity,

\begin{subequations}
\begin{align}
\Psi &= e^{-i \omega r_{*}} + R e^{i \omega r_{*}},  r_{*} \to \infty, \\ 
\Psi &= T e^{-i \omega r_{*}},   r_{*} \to -\infty, 
\end{align}
\end{subequations}

where R and T are the reflection and transmission coefficients.
This scenario is equivalent to the scattering of a wave 
originating from the horizon.

The nature of the effective potential is such that it decreases at both
infinity, therefore we can safely apply the WKB method 
\cite{Schutz:1985km,Iyer:1986np,Konoplya:2003ii,Konoplya:2004ip} to find 
reflection and transmission coefficients. The reflection and 
transmission coefficients satisfy the following relations

\begin{equation}
    |R|^2 +|T|^2=1.
\end{equation}

Once the reflection coefficient is computed, we can find the 
transmission coefficient for each multipole number as follows

\begin{equation}
    |A_{l}|^2 =1- |R_{l}|^2.
\end{equation}

Numerous approaches are available in the literature for the computation of the transmission 
and reflection coefficients. For an accurate computation of the transmission 
and reflection coefficients, we employed the 6th-order WKB formula 
\cite{Schutz:1985km,Iyer:1986np,Konoplya:2003ii,Konoplya:2004ip}. In accordance 
with the findings presented in 
\cite{Schutz:1985km,Iyer:1986np,Konoplya:2003ii,Konoplya:2004ip}, the reflection 
coefficient can be written as 

\begin{equation}
    R= (1+e^{-2i\pi K})^{-\frac{1}{2}},
\end{equation}

where $K$ is defined by the following relations

\begin{equation}
  K - i \frac{(\omega^2 - V_0)}{\sqrt{-2V_{0}^{\prime \prime}}} - \sum_{i=2}^{i=6} \Lambda_{i}(K)=0,
\end{equation}

where $V_{0}$ is the effective potential at $r=r_{max}$, $V_{0}^{\prime \prime}$ is 
the double derivative of effective potential with respect to radial coordinate $r$, 
evaluated at $r=r_{max}$ and $\Lambda_i$ is the higher-order correction to the WKB
formula.

\subsection{Quadratic GB-Coupling Functional}\label{sec:4.1}

\begin{figure}[H]
\centering
\subfloat[$l=1$ \& $s=1$]{\includegraphics[width=.5\textwidth]{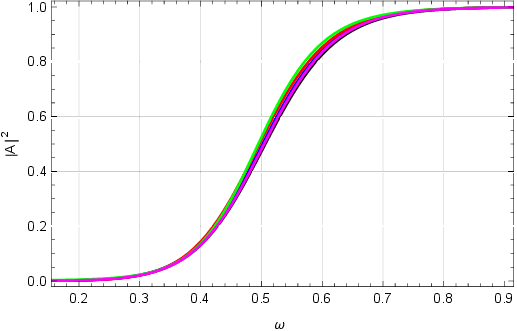}}\hfill
\subfloat[$l=2$ \& $s=1$]{\includegraphics[width=.5\textwidth]{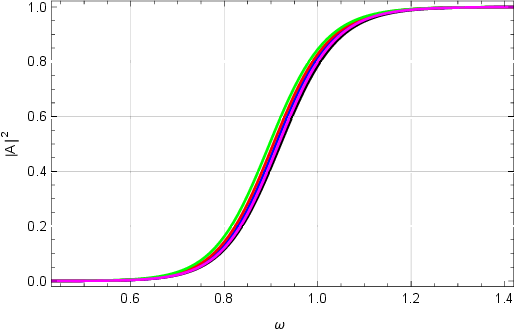}}\hfill
\subfloat[$l=3$ \& $s=1$]{\includegraphics[width=.5\textwidth]{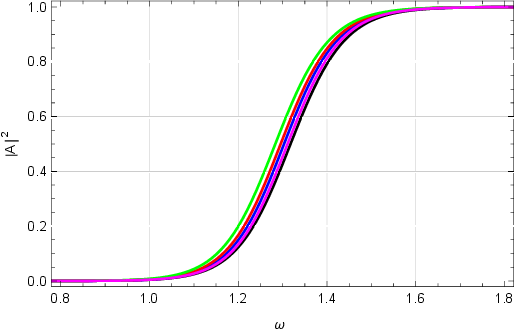}}\hfill
\caption{The dependence of grey-body factor on $\omega$, for the 
electromagnetic field $(s = 1)$, with $r_0=1$, black line denotes $p=0.0$, 
magenta line denotes $p=0.2$, blue line denotes $p=0.4$, red line denotes $p=0.6$ 
and green line denotes $p=0.8$.}\label{fig:46}
\end{figure}

The grey--body factors for quadratic GB-coupling functional are shown in Fig. \ref{fig:46} as a function of $\omega$ for electromagnetic fields with different values of multiple numbers $l$. From Fig. \ref{fig:46} it can be seen that the grey--body factor is higher for EsGB black holes compared to Schwarzschild black holes. As the parameter $p$ increases transmission rate of the particles also increases, which is consistent with the nature of the effective potential, i.e., as the parameter $p$ increases the height of the potential barrier becomes lower resulting in a high transmission rate for the particles and vice-versa.

\subsection{Cubic GB-Coupling Functional}\label{sec:4.2}

\begin{figure}[H]
\centering
\subfloat[$l=1$ \& $s=1$]{\includegraphics[width=.5\textwidth]{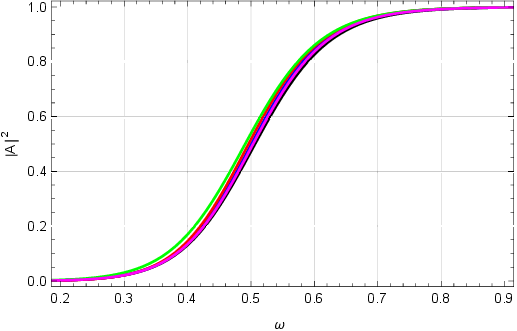}}\hfill
\subfloat[$l=2$ \& $s=1$]{\includegraphics[width=.5\textwidth]{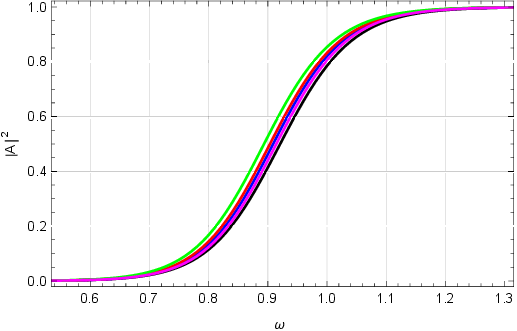}}\hfill
\subfloat[$l=3$ \& $s=1$]{\includegraphics[width=.5\textwidth]{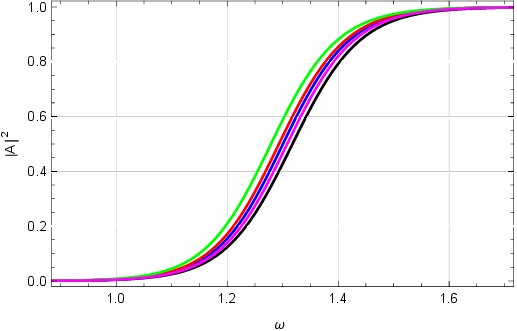}}\hfill
\caption{The dependence of grey-body factor on $\omega$, for the electromagnetic 
field $(s = 1)$, with $r_0=1$, black line denotes $p=0.0$, magenta line denotes 
$p=0.2$, blue line denotes $p=0.4$, red line denotes $p=0.6$ and green line 
denotes $p=0.8$.}\label{fig:47}
\end{figure}

The grey--body factors for cubic GB-coupling functional are shown in Fig. \ref{fig:47} as a function of $\omega$ for electromagnetic fields with different values of multiple numbers $l$. From Fig. \ref{fig:47} it can be seen that the grey--body factor is higher for EsGB black holes compared to Schwarzschild black holes. As the parameter $p$ increases transmission rate of the particles also increases, which is consistent with the nature of the effective potential, i.e., as the parameter $p$ increases the height of the potential barrier becomes lower resulting in a high transmission rate for the particles and vice-versa.

\subsection{Quartic GB-Coupling Functional}\label{sec:4.3}

\begin{figure}[H]
\centering
\subfloat[$l=1$ \& $s=1$]{\includegraphics[width=.5\textwidth]{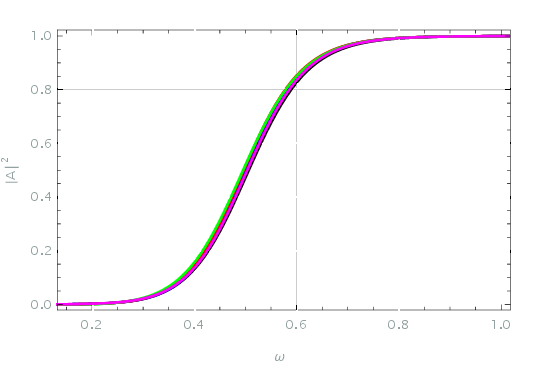}}\hfill
\subfloat[$l=2$ \& $s=1$]{\includegraphics[width=.5\textwidth]{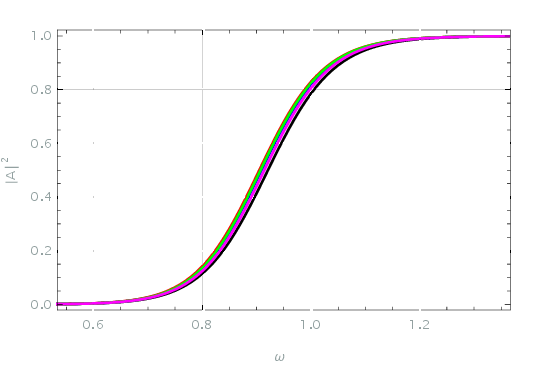}}\hfill
\subfloat[$l=3$ \& $s=1$]{\includegraphics[width=.5\textwidth]{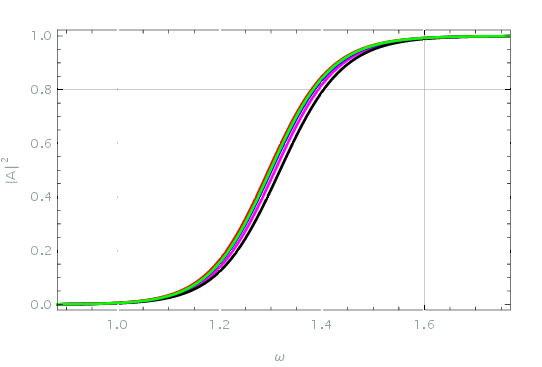}}\hfill
\caption{The dependence of grey-body factor on $\omega$, for the 
electromagnetic field $(s = 1)$, with $r_0=1$, black line denotes 
$p=0.0$, magenta line denotes $p=0.2$, blue line denotes $p=0.4$, 
red line denotes $p=0.6$ and green line denotes $p=0.8$.}\label{fig:48}
\end{figure}

The grey--body factors for quartic GB-coupling functional are shown in Fig. \ref{fig:48} as a function of $\omega$ for electromagnetic fields with different values of multiple numbers $l$. From Fig. \ref{fig:48} it can be seen that the grey--body factor is higher for EsGB black holes compared to Schwarzschild black holes. As the parameter $p$ increases transmission rate of the particles also increases, which is consistent with the nature of the effective potential, i.e., as the parameter $p$ increases the height of the potential barrier becomes lower resulting in a high transmission rate for the particles and vice-versa.

\subsection{Inverse GB-Coupling Functional}\label{sec:4.4}

\begin{figure}[H]
\centering
\subfloat[$l=1$ \& $s=1$]{\includegraphics[width=.5\textwidth]{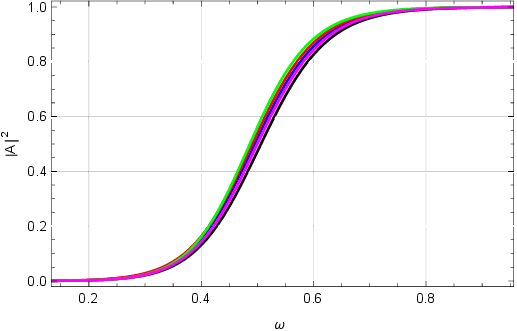}}\hfill
\subfloat[$l=2$ \& $s=1$]{\includegraphics[width=.5\textwidth]{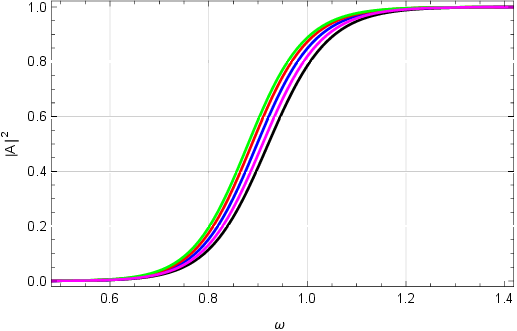}}\hfill
\subfloat[$l=3$ \& $s=1$]{\includegraphics[width=.5\textwidth]{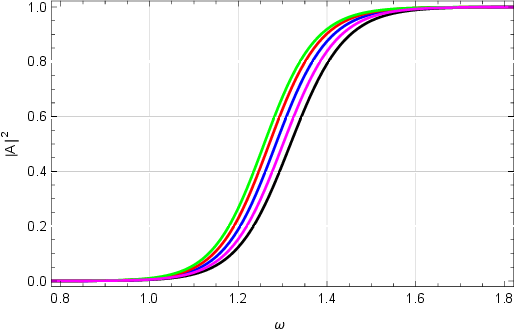}}\hfill
\caption{The dependence of grey-body factor on $\omega$, for the 
electromagnetic field $(s = 1)$, with $r_0=1$, black line denotes $p=0.0$, 
magenta line denotes $p=0.2$, blue line denotes $p=0.4$, red line denotes 
$p=0.6$ and green line denotes $p=0.8$.}\label{fig:49}
\end{figure}

The grey--body factors for inverse GB-coupling functional are shown in Fig. \ref{fig:49} as a function of $\omega$ for electromagnetic fields with different values of multiple numbers $l$. From Fig. \ref{fig:49} it can be seen that the grey--body factor is higher for EsGB black holes compared to Schwarzschild black holes. As the parameter $p$ increases transmission rate of the particles also increases, which is consistent with the nature of the effective potential, i.e., as the parameter $p$ increases the height of the potential barrier becomes lower resulting in a high transmission rate for the particles and vice-versa.

\subsection{Logarithmic GB-Coupling Functional}\label{sec:4.5}

\begin{figure}[H]
\centering
\subfloat[$l=1$ \& $s=1$]{\includegraphics[width=.5\textwidth]{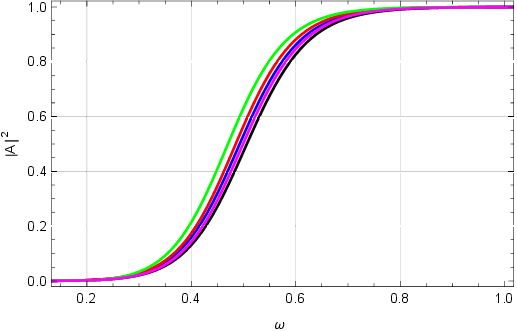}}\hfill
\subfloat[$l=2$ \& $s=1$]{\includegraphics[width=.5\textwidth]{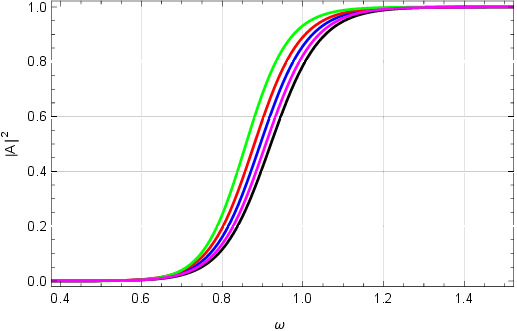}}\hfill
\subfloat[$l=3$ \& $s=1$]{\includegraphics[width=.5\textwidth]{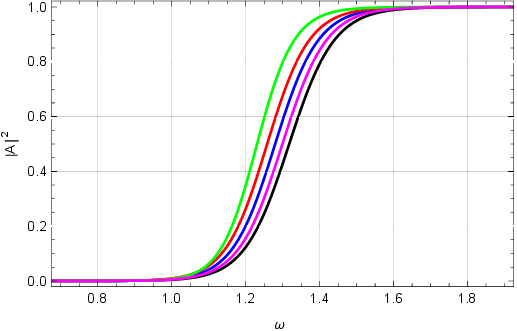}}\hfill
\caption{The dependence of grey-body factor on $\omega$, for 
the electromagnetic field $(s = 1)$, with $r_0=1$, black line denotes 
$p=0.0$, magenta line denotes $p=0.2$, blue line denotes $p=0.4$, red 
line denotes $p=0.6$ and green line denotes $p=0.8$.}\label{fig:50}
\end{figure}

The grey--body factors for logarithmic GB-coupling functional are shown in Fig. \ref{fig:50} as a function of $\omega$ for electromagnetic fields with different values of multiple numbers $l$. From Fig. \ref{fig:50} it can be seen that the grey--body factor is higher for EsGB black holes compared to Schwarzschild black holes. As the parameter $p$ increases transmission rate of the particles also increases, which is consistent with the nature of the effective potential, i.e., as the parameter $p$ increases the height of the potential barrier becomes lower resulting in a high transmission rate for the particles and vice-versa.

Here, we will write a brief analysis of how different GB-coupling functions affect the quasinormal modes of the black hole. From above analysis of the quasinormal modes of EsGB black hole as a function of parameter $p$, we see a strong deviation of qnms for both massless scalar and electromagnetic field compared to Schwarzschild black hole when GB-coupling functions are logarithmic. The deviation of quasinormal modes is lowest for quadratic GB-coupling functional. The transmission rate of the particles is highest for logarithmic GB-coupling functional (i.e., the strongest deviation of transmission rate occurs compared to Schwarzschild black hole) and lowest for quadratic GB-coupling functional (i.e., the lowest deviation of transmission rate occurs compared to Schwarzschild black hole).

\section{Conclusions}\label{sec:5}
In this study, we address previously unexplored aspects of quasinormal modes 
for scalar and electromagnetic fields in the vicinity of analytically obtained 
EsGB black holes solutions. Here, we investigate:
\begin{itemize}
\item The quasinormal modes of scalar and electromagnetic fields in the background 
    of EsGB black hole for different GB coupling functional (quadratic, cubic, 
    quartic, inverse, and Logarithmic). The dependence of the real and imaginary parts 
    of the quasinormal modes on the parameter $p$ is shown in Figures. For $p=0$, 
    quasinormal modes of EsGB black hole match with Schwarzschild one. For non--zero 
    values of $p$ quasinormal frequency of EsGB black hole take smaller values compared 
    to Schwarzschild black hole. 
        
\item We analyzed massive scalar fields in the EsGB background. The 
height of the effective potential for massive scalar fields is higher as the mass 
increases. For large values of mass, it shows undamped oscillations occur, which 
commonly known as quasi--resonances. 

\item We derived an analytical formula of quasinormal frequency in the eikonal 
regime for each GB coupling function. In the limit $p \to 0$, 
it reduced to the quasinormal frequency of Schwarzschild black hole. We have shown that the correspondence between the eikonal quasinormal frequencies and null geodesics \cite{Cardoso:2008bp} is hold in the EsGB theory for the test fields under consideration. 

\item Finally, we studied the grey-body factors for electromagnetic fields with 
different values of multiple numbers. Our analysis showed that for higher values 
of parameter $p$ grey-body factors are larger.

\end{itemize}

\section{Acknowledgements}
I would like to thank Roman Konoplya for useful discussions and help with the Mathematica WKB code.

\begin{appendices}
\setcounter{equation}{0}
\renewcommand\theequation{\thesection\arabic{equation}} 
\section{Appendix: Analytical expressions for the metric functions up to fourth order}\label{sec:6}
Here, we will write down the analytical expression for the metric functions
up to the fourth order in the CFA method. Therefore,
the metric functions are given by

\begin{equation}
   g_{tt}(r) \approx \frac{\mathcal{N}^1}{\mathcal{D}^1}(1- \frac{r_0}{r}),
\end{equation}
\begin{equation}
    \sqrt{g_{tt}(r)g_{rr}(r)} \approx \frac{\mathcal{N}^2}{\mathcal{D}^2},
\end{equation}
where  the numerators, $\mathcal{N}^1$, $\mathcal{N}^2$ and denominators,
$\mathcal{D}^1$, $\mathcal{D}^2$ are separately given for each GB coupling 
functional.

\subsection{Quadratic GB-Coupling Functional: \texorpdfstring{$f(\varphi)=\varphi^2$}{TEXT}}

\begin{equation}
    g_{tt}(r) \approx \frac{\mathcal{N}^1_{eve2}}{\mathcal{D}^1_{eve2}}(1- \frac{r_0}{r}),
 \end{equation}
 \begin{equation}
     \sqrt{g_{tt}(r)g_{rr}(r)} \approx \frac{\mathcal{N}^2_{eve2}}{\mathcal{D}^2_{eve2}},
 \end{equation}
 where  
\begin{equation}
\begin{aligned}   
\mathcal{N}^1_{\text{eve}2}  & = p^{11} (-0.027339 r^3 r_{0}+0.027339r^2r_{0}^2+0.027339 r {r_{0}}^3-0.027339 {r_{0}}^4)+p^{10}(-1.65064r^3{r_{0}} \\
                             & +r^4+0.639158 r^2 {r_{0}}^2 -0.349358 r {r_{0}}^3+0.360842 {r_{0}}^4)+p^9 (-13.5993 r^4+25.9764 r^3 {r_{0}} \\
                             &-12.2194 r^2 {r_{0}}^2+1.96871r{r_{0}}^3-2.12636 {r_{0}}^4)+p^8 (66.7223 r^4-136.15 r^3 {r_{0}}+68.5796 r^2 {r_{0}}^2 \\
                             & -6.96288 r {r_{0}}^3+7.81103 {r_{0}}^4) +p^7 (-142.79 r^4+317.427 r^3 {r_{0}}-172.292 r^2 {r_{0}}^2+15.8938 r{r_{0}}^3 \\
                             & -18.2388 {r_{0}}^4)+p^6 (101.122 r^4-301.89 r^3 {r_{0}} +197.051 r^2 {r_{0}}^2 -22.9918 r {r_{0}}^3+26.7025 {r_{0}}^4) \\
                             & +p^5 (120.603 r^4-78.7254 r^3 {r_{0}}-38.3503 r^2 {r_{0}}^2+20.5997 r {r_{0}}^3-24.0645 {r_{0}}^4) +p^4 (-295.148 r^4 \\
                             & +437.019 r^3 {r_{0}}-143.957 r^2 {r_{0}}^2-10.8823 r {r_{0}}^3+12.7195 {r_{0}}^4) +p^3 (236.149 r^4-388.645 r^3 {r_{0}} \\
                             & +153.456 r^2 {r_{0}}^2+3.01323 r {r_{0}}^3-3.47898 {r_{0}}^4) +p^2 (-85.9513 r^4 +147.171 r^3 {r_{0}}-61.7593 r^2 {r_{0}}^2 \\
                             & -0.319982 r {r_{0}}^3 +0.336681 {r_{0}}^4)+p (12.5012 r^4-21.3996 r^3 {r_{0}}+9.1727 r^2 {r_{0}}^2+0.00359074 r {r_{0}}^3 \\
                             & +0.0054195 {r_{0}}^4) -0.608781 r^4 +0.894103 r^3 {r_{0}} -0.347264 r^2 {r_{0}}^2, 
\end{aligned}
\end{equation}

\begin{equation}
    \begin{aligned}
\mathcal{D}^1_{\text{eve}2} & = p^{10} r^2 ( r^2-2 r {r_{0}}+ {r_{0}}^2)+p^9 r^2 (-13.5993 r^2+27.6187 r {r_{0}}-14.0194 {r_{0}}^2)+p^8 r^2 (66.7223 r^2 \\
                            & -139.556 r {r_{0}}+72.8336 {r_{0}}^2)+p^7 r^2 (-142.79 r^2+319.724 r {r_{0}}-176.934 {r_{0}}^2)+p^6 r^2 (101.122 r^2 \\
                            & -298.896 r {r_{0}}+197.772 {r_{0}}^2)+p^5 r^2 (120.603 r^2-85.7535 r {r_{0}}-34.8151 {r_{0}}^2)+p^4 r^2 (-295.148 r^2 \\
                            & +442.573 r {r_{0}}-147.595 {r_{0}}^2)+p^3 r^2 (236.149 r^2-390.653 r {r_{0}}+154.893 {r_{0}}^2)+p^2 r^2 (-85.9513 r^2\\
                            & +147.462 r {r_{0}}-61.9667 {r_{0}}^2)+p r^2 (12.5012 r^2-21.4138 r {r_{0}}+9.17934 {r_{0}}^2)+r^2 (-0.608781 r^2\\
                            & +0.894103 r {r_{0}}-0.347264 {r_{0}}^2),
    \end{aligned}
\end{equation}

\begin{equation}
    \begin{aligned}
\mathcal{N}^2_{\text{eve}2} & = p^9 ( r^3-2 r^2 {r_{0}}+ r {r_{0}}^2)+p^8 (-8.86767 r^3+17.7041 r^2 {r_{0}}-8.8364 r {r_{0}}^2)+p^7 (29.1248 r^3 \\
                            & -57.7325 r^2 {r_{0}}+28.8414 r {r_{0}}^2-0.233666 {r_{0}}^3)+p^6 (-56.6765 r^3+110.257 r^2 {r_{0}}-54.735 r {r_{0}}^2 \\
                            & +1.1405 {r_{0}}^3)+p^5 (90.3253 r^3-171.817 r^2 {r_{0}}+83.6835 r {r_{0}}^2-2.08071 {r_{0}}^3)+p^4 (-119.877 r^3 \\
                            & +226.049 r^2 {r_{0}}-108.183 r {r_{0}}^2+1.65941 {r_{0}}^3)+p^3 (105.82 r^3-199.251 r^2 {r_{0}}+94.4276 r {r_{0}}^2 \\
                            & -0.411268 {r_{0}}^3)+p^2 (-51.741 r^3+96.6288 r^2 {r_{0}}-45.2948 r {r_{0}}^2-0.132824 {r_{0}}^3)+p (12.1153 r^3 \\
                            & -21.8254 r^2 {r_{0}}+9.91398 r {r_{0}}^2+0.058561 {r_{0}}^3)-1.22281 r^3+1.98757 r^2 {r_{0}}-0.817317 r {r_{0}}^2,
    \end{aligned}
\end{equation}

\begin{equation}
    \begin{aligned}
\mathcal{D}^2_{\text{eve}2} & = p^9 r (1. r^2-2. r {r_{0}}+1. {r_{0}}^2)+p^8 r (-8.86767 r^2+17.7041 r {r_{0}}-8.8364 {r_{0}}^2)+p^7 r (29.1248 r^2 \\
                            & -57.7325 r {r_{0}}+28.6077 {r_{0}}^2)+p^6 r (-56.6765 r^2+110.257 r {r_{0}}-53.5872 {r_{0}}^2)+p^5 r (90.3253 r^2 \\
                            & -171.817 r {r_{0}}+81.5611 {r_{0}}^2)+p^4 r (-119.877 r^2+226.049 r {r_{0}}-106.429{r_{0}}^2)+p^3 r (105.82 r^2 \\
                            &-199.251 r {r_{0}}+93.9092 {r_{0}}^2)+p^2 r (-51.741 r^2+96.6288 r {r_{0}}-45.3674 {r_{0}}^2)+p r (12.1153 r^2 \\
                            & -21.8254 r {r_{0}}+9.95903 {r_{0}}^2)+r (-1.22281 r^2+1.98757 r {r_{0}}-0.817317 {r_{0}}^2).
    \end{aligned}
\end{equation}
\subsection{Cubic GB-Coupling Functional: \texorpdfstring{$f(\varphi)=\varphi^3$}{TEXT}}

\begin{equation}
    g_{tt}(r) \approx \frac{\mathcal{N}^1_{odd3}}{\mathcal{D}^1_{odd3}}(1- \frac{r_0}{r}),
 \end{equation}
 \begin{equation}
     \sqrt{g_{tt}(r)g_{rr}(r)} \approx \frac{\mathcal{N}^2_{odd3}}{\mathcal{D}^2_{odd3}},
 \end{equation}

 where 
 \begin{equation}
    \begin{aligned}   
    \mathcal{N}^1_{\text{odd}3}  & = p^{12} (0.00537634 r^3 r_{0}-0.00537634 r^2 r_{0}^2-0.00537634 r r_{0}^3+0.00537634 r_{0}^4) \\
                                 & +p^{11} (-0.0670504 r^3 r_{0}+0.0683756 r^2 r_{0}^2+0.0670504 r r_{0}^3-0.0683756 r_{0}^4)+p^{10} (1. r^4 \\
                                 & -1.70036 r^3 r_{0}+0.683358 r^2 r_{0}^2-0.299635 r r_{0}^3+0.316642 r_{0}^4)+p^9 (-6.78182 r^4+13.1727 r^3 r_{0} \\
                                 & -6.31042 r^2 r_{0}^2+0.737004 r r_{0}^3-0.817608 r_{0}^4)+p^8 (17.7971 r^4-36.7027 r^3 r_{0}+18.714 r^2 r_{0}^2 \\
                                 & -1.12269 r r_{0}^3+1.3156 r_{0}^4)+p^7 (-21.7877 r^4+48.5071 r^3 r_{0}-26.4795 r^2 r_{0}^2+1.08356 r r_{0}^3 \\
                                 & -1.34315 r_{0}^4)+p^6 (9.51766 r^4-26.9583 r^3 r_{0} +17.3344 r^2 r_{0}^2-0.638392 r r_{0}^3+0.834102 r_{0}^4) \\
                                 & +p^5 (4.59852 r^4-2.91241 r^3 r_{0}-1.79709 r^2 r_{0}^2 +0.19656 r r_{0}^3-0.266067 r_{0}^4)+p^4 (-5.37404 r^4 \\
                                 & +8.71817 r^3 r_{0}-3.1644 r^2 r_{0}^2-0.0116997 r r_{0}^3 +0.0128034 r_{0}^4)+p^3 (0.540319 r^4-1.28023 r^3 r_{0} \\
                                 & +0.652328 r^2 r_{0}^2-0.00618425 r r_{0}^3+0.0102582 r_{0}^4)+p^2 (0.451911 r^4-0.727369 r^3 r_{0}+0.284135 r^2 r_{0}^2 \\
                                 & -0.00017735 r r_{0}^3+0.000404216 r_{0}^4)+p (0.0379465 r^4-0.0549725 r^3 r_{0}+0.0202381 r^2 r_{0}^2-0.0000201357 r r_{0}^3 \\
                                 & +0.0000179163 r_{0}^4)+0.000140052 r^4-0.0000971891 r^3 r_{0},
    \end{aligned}
    \end{equation}
    
    \begin{equation}
        \begin{aligned}
    \mathcal{D}^1_{\text{odd}3} & = p^{10} r^2 (1. r^2-2. r r_{0}+1. r_{0}^2)+p^9 r^2 (-6.78182 r^2+13.8101 r r_{0}-7.02829 r_{0}^2) \\
                                & +p^8 r^2 (17.7971 r^2-37.355 r r_{0}+19.5579 r_{0}^2)+p^7 r^2 (-21.7877 r^2+48.6892 r r_{0}-26.9117 r_{0}^2) \\
                                & +p^6 r^2 (9.51766 r^2 -26.7448 r r_{0}+17.2901 r_{0}^2)+p^5 r^2 (4.59852 r^2-3.06912 r r_{0}-1.67657 r_{0}^2) \\
                                & +p^4 r^2 (-5.37404 r^2+8.71215 r r_{0} -3.17516 r_{0}^2)+p^3 r^2 (0.540319 r^2-1.2622 r r_{0}+0.6403 r_{0}^2) \\
                                & +p^2 r^2 (0.451911 r^2-0.725183 r r_{0}+0.28313 r_{0}^2)+p r^2 (0.0379465 r^2-0.0549644 r r_{0}+0.0202407 r_{0}^2) \\
                                & +r^2 (0.000140052 r^2-0.0000971891 r r_{0}),
        \end{aligned}
    \end{equation}

    \begin{equation}
        \begin{aligned}
    \mathcal{N}^2_{\text{odd}3} & = p^9 (1. r^3-2. r^2 r_{0}+1. r r_{0}^2)+p^8 (-15.624 r^3+31.3197 r^2 r_{0}-15.587 r r_{0}^2-0.108754 r_{0}^3)+p^7 (75.8827 r^3 \\
                                & -153.129 r^2 r_{0}+75.7277 r r_{0}^2+1.51139 r_{0}^3)+p^6 (-135.885 r^3+279.413 r^2 r_{0}-138.956 r r_{0}^2-4.46757 r_{0}^3) \\
                                & +p^5 (72.0953 r^3-161.504 r^2 r_{0}+83.7029 r r_{0}^2+5.22405 r_{0}^3)+p^4 (47.4153 r^3-75.3996 r^2 r_{0}+31.4976 r r_{0}^2 \\
                                & -2.47728 r_{0}^3)+p^3 (-61.6661 r^3+111.092 r^2 r_{0}-50.8295 r r_{0}^2+0.253284 r_{0}^3)+p^2 (18.923 r^3-32.9427 r^2 r_{0} \\
                                & +14.5941 r r_{0}^2+0.0638083 r_{0}^3)+p (-2.11497 r^3+3.11041 r^2 r_{0}-1.13326 r r_{0}^2+0.0010564 r_{0}^3)-0.0225954 r^3 \\
                                & +0.0327384 r^2 r_{0}-0.0128335 r r_{0}^2,
        \end{aligned}
    \end{equation}

    \begin{equation}
        \begin{aligned}
    \mathcal{D}^2_{\text{odd}3} & = p^9 r (1. r^2-2. r r_{0}+1. r_{0}^2)+p^8 r (-15.624 r^2+31.3197 r r_{0}-15.6957 r_{0}^2)+p^7 r (75.8827 r^2 \\
                                & -153.129 r r_{0}+77.2427 r_{0}^2)+p^6 r (-135.885 r^2+279.413 r r_{0}-143.469 r_{0}^2)+p^5 r (72.0953 r^2-161.504 r r_{0} \\
                                & +89.0785 r_{0}^2)+p^4 r (47.4153 r^2-75.3996 r r_{0}+28.8021 r_{0}^2)+p^3 r (-61.6661 r^2+111.092 r r_{0}-50.4318 r_{0}^2) \\
                                & +p^2 r (18.923 r^2 -32.9427 r r_{0}+14.6226 r_{0}^2)+p r (-2.11497 r^2+3.11041 r r_{0}-1.13289 r_{0}^2)+r (-0.0225954 r^2 \\
                                & +0.0327384 r r_{0}-0.0128335 r_{0}^2).
        \end{aligned}
    \end{equation}

\subsection{Quartic GB-Coupling Functional: \texorpdfstring{$f(\varphi)=\varphi^4$}{TEXT}}

\begin{equation}
        g_{tt}(r) \approx \frac{\mathcal{N}^1_{eve4}}{\mathcal{D}^1_{eve4}}(1- \frac{r_0}{r}),
\end{equation}
\begin{equation}
         \sqrt{g_{tt}(r)g_{rr}(r)} \approx \frac{\mathcal{N}^2_{eve4}}{\mathcal{D}^2_{eve4}},
\end{equation}
    
where     
\begin{equation}
 \begin{aligned}   
    \mathcal{N}^1_{\text{eve}4}  & = p^{15} (1. r^4-2.089 r^3 r_{0}+1.089 r^2 r_{0}^2+0.0889959 r r_{0}^3-0.0889959 r_{0}^4)+p^{14} (-6.89482 r^4 \\
                                 & +14.7896 r^3 r_{0}-7.90886 r^2 r_{0}^2-0.757057 r r_{0}^3+0.771153 r_{0}^4)+p^{13} (12.5106 r^4-29.4208 r^3 r_{0} \\
                                 & +17.0451 r^2 r_{0}^2+2.92063 r r_{0}^3-3.05555 r_{0}^4)+p^{12} (19.4917 r^4-29.885 r^3 r_{0}+9.84773 r^2 r_{0}^2 \\
                                 & -6.69766 r r_{0}^3+7.24323 r_{0}^4)+p^{11} (-115.113 r^4+224.275 r^3 r_{0}-107.925 r^2 r_{0}^2+9.97888 r r_{0}^3 \\
                                 & -11.206 r_{0}^4)+p^{10} (198.927 r^4-410.675 r^3 r_{0}+210.013 r^2 r_{0}^2-9.86857 r r_{0}^3+11.562 r_{0}^4) \\
                                 & +p^9 (-164.194 r^4+364.121 r^3 r_{0}-198.46 r^2 r_{0}^2+6.32619 r r_{0}^3-7.79975 r_{0}^4)+p^8 (45.2779 r^4 \\
                                 & -130.919 r^3 r_{0}+85.2 r^2 r_{0}^2-2.38976 r r_{0}^3+3.15916 r_{0}^4)+p^7 (28.2324 r^4-33.1842 r^3 r_{0} \\
                                 & +4.40023 r^2 r_{0}^2+0.342238 r r_{0}^3-0.525779 r_{0}^4)+p^6 (-21.3691 r^4+38.4559 r^3 r_{0}-16.3563 r^2 r_{0}^2 \\
                                 & +0.0898093 r r_{0}^3-0.104246 r_{0}^4)+p^5 (-0.197567 r^4-1.57336 r^3 r_{0}+1.45736 r^2 r_{0}^2-0.0324933 r r_{0}^3 \\
                                 & +0.0403357 r_{0}^4)+p^4 (2.04375 r^4-3.45917 r^3 r_{0}+1.43513 r^2 r_{0}^2-0.00124132 r r_{0}^3+0.00429355 r_{0}^4) \\
                                 & +p^3 (0.276656 r^4-0.423294 r^3 r_{0}+0.159108 r^2 r_{0}^2+0.0000460001 r r_{0}^3+0.000142526 r_{0}^4)+p^2 (0.00912579 r^4 \\
                                 & -0.0124531 r^3 r_{0}+0.0040693 r^2 r_{0}^2-5.79588 \times 10^{-6} r r_{0}^3+5.6289 \times 10^{-6} r_{0}^4) \\
                                 & +p (-0.0000509404 r^4+0.000117355 r^3 r_{0}-0.0000607436 r^2 r_{0}^2+4.44458 \times 10^{-8} r r_{0}^3 \\
                                 & -6.14644 \times 10^{-8} r_{0}^4)-2.63557 \times 10^{-7} r^4+1.63201 \times 10^{-8} r^3 r_{0}+ 1.79379 \times 10^{-7} r^2 r_{0}^2,
 \end{aligned}
\end{equation}
        
\begin{equation}
\begin{aligned}
    \mathcal{D}^1_{\text{eve}4} & = p^{15} r^2 (1. r^2-2. r r_{0}+1. r_{0}^2)+p^{14} r^2 (-6.89482 r^2+13.948 r r_{0}-7.05321 r_{0}^2)+p^{13} r^2 (12.5106 r^2 \\
                                & -26.1315 r r_{0}+13.6209 r_{0}^2)+p^{12} r^2 (19.4917 r^2-36.7725 r r_{0}+17.2808 r_{0}^2)+p^{11} r^2 (-115.113 r^2 \\
                                & +232.465 r r_{0}-117.345 r_{0}^2)+p^{10} r^2 (198.927 r^2-415.683 r r_{0}+216.716 r_{0}^2)+p^9 r^2 (-164.194 r^2+364.547 r r_{0} \\
                                & -200.316 r_{0}^2)+p^8 r^2 (45.2779 r^2-129.549 r r_{0}+84.468 r_{0}^2)+p^7 r^2 (28.2324 r^2-33.7905 r r_{0}+4.977 r_{0}^2) \\
                                & +p^6 r^2 (-21.3691 r^2+38.3498 r r_{0}-16.341 r_{0}^2)+p^5 r^2 (-0.197567 r^2-1.50422 r r_{0}+1.40334 r_{0}^2) \\
                                & +p^4 r^2 (2.04375 r^2-3.44431 r r_{0}+1.42689 r_{0}^2)+p^3 r^2 (0.276656 r^2-0.422662 r r_{0}+0.158858 r_{0}^2) \\
                                & +p^2 r^2 (0.00912579 r^2-0.0124564 r r_{0}+0.00407385 r_{0}^2)+p r^2 (-0.0000509404 r^2+0.000117338 r r_{0} \\
                                & -0.0000607602 r_{0}^2) +r^2 (-2.63557 \times 10^{-7} r^2+1.63201 \times 10^{-8} r r_{0}+1.79379 \times 10^{-7} r_{0}^2),
\end{aligned}
\end{equation}

\begin{equation}
\begin{aligned}
    \mathcal{N}^2_{\text{eve}4} & = p^{10} (1. r^3-2. r^2 r_{0}+1. r r_{0}^2)+p^9 (-13.6797 r^3+27.4429 r^2 r_{0}-13.6766 r r_{0}^2-0.0865879 r_{0}^3) \\
                                & +p^8 (67.4124 r^3-135.988 r^2 r_{0}+67.7236 r r_{0}^2+0.852114 r_{0}^3)+p^7 (-148.278 r^3+302.545 r^2 r_{0}-151.49 r r_{0}^2 \\
                                & -2.77603 r_{0}^3)+p^6 (146.99 r^3-308.586 r^2 r_{0}+157.352 r r_{0}^2+4.22482 r_{0}^3)+p^5 (-36.5144 r^3+91.761 r^2 r_{0} \\
                                & -51.9412 r r_{0}^2-3.22666 r_{0}^3)+p^4 (-41.1576 r^3+69.4003 r^2 r_{0}-29.5381 r r_{0}^2+1.13594 r_{0}^3)+p^3 (28.9496 r^3 \\
                                & -53.2293 r^2 r_{0}+24.5671 r r_{0}^2-0.117809 r_{0}^3)+p^2 (-5.11947 r^3+9.28066 r^2 r_{0}-4.24613 r r_{0}^2-0.00574379 r_{0}^3) \\
                                & +p (0.391241 r^3-0.618344 r^2 r_{0}+0.245794 r r_{0}^2-0.0000557057 r_{0}^3)+0.00525892 r^3-0.00809504 r^2 r_{0} \\
                                & +0.00330483 r r_{0}^2,
\end{aligned}
\end{equation}

\begin{equation}
\begin{aligned}
    \mathcal{D}^2_{\text{eve}4} & = p^{10} r (1. r^2-2. r r_{0}+1. r_{0}^2)+p^9 r (-13.6797 r^2+27.4429 r r_{0}-13.7632 r_{0}^2)+p^8 r (67.4124 r^2 \\
                                & -135.988 r r_{0}+68.5757 r_{0}^2)+p^7 r (-148.278 r^2+302.545 r r_{0}-154.267 r_{0}^2)+p^6 r (146.99 r^2-308.586 r r_{0} \\
                                & +161.585 r_{0}^2)+p^5 r (-36.5144 r^2+91.761 r r_{0}-55.1917 r_{0}^2)+p^4 r (-41.1576 r^2+69.4003 r r_{0}-28.3697 r_{0}^2) \\
                                & +p^3 r (28.9496 r^2-53.2293 r r_{0}+24.4286 r_{0}^2)+p^2 r (-5.11947 r^2+9.28066 r r_{0}-4.24704 r_{0}^2)+p r (0.391241 r^2 \\
                                & -0.618344 r r_{0}+0.245858 r_{0}^2)+r (0.00525892 r^2-0.00809504 r r_{0}+0.00330483 r_{0}^2).
\end{aligned}
\end{equation}

\subsection{Inverse GB-Coupling Functional: \texorpdfstring{$f(\varphi)=\varphi^{-1}$}{TEXT}}

\begin{equation}
        g_{tt}(r) \approx \frac{\mathcal{N}^1_{inv}}{\mathcal{D}^1_{inv}}(1- \frac{r_0}{r}),
\end{equation}
\begin{equation}
         \sqrt{g_{tt}(r)g_{rr}(r)} \approx \frac{\mathcal{N}^2_{inv}}{\mathcal{D}^2_{inv}},
\end{equation}
    
where     
\begin{equation}
 \begin{aligned}   
    \mathcal{N}^1_{\text{inv}}  & = p^{13} (0.00630915 r^3 {r_{0}}-0.00630915 r^2 {r_{0}}^2-0.00630915 r {r_{0}}^3+0.00630915 {r_{0}}^4) \\
                                & +p^{12} (-0.288247 r^3 {r_{0}}+0.289835 r^2 {r_{0}}^2+0.288247 r {r_{0}}^3-0.289835 {r_{0}}^4) \\
                                & +p^{11} (1. r^4+1.44067 r^3 {r_{0}}-2.51152 r^2 {r_{0}}^2-2.56824 r {r_{0}}^3+2.6391 {r_{0}}^4) \\
                                & +p^{10} (-33.5247 r^4+55.6895 r^3 {r_{0}}-21.3738 r^2 {r_{0}}^2+8.14436 r {r_{0}}^3-8.93465 {r_{0}}^4) \\
                                & +p^9 (139.634 r^4-272.59 r^3 {r_{0}}+130.873 r^2 {r_{0}}^2-10.712 r {r_{0}}^3+12.9426 {r_{0}}^4) \\
                                & +p^8 (-211.343 r^4+445.224 r^3 {r_{0}}-232.351 r^2 {r_{0}}^2+3.40437 r {r_{0}}^3-5.11853 {r_{0}}^4) \\
                                & +p^7 (104.307 r^4-239.683 r^3 {r_{0}}+134.352 r^2 {r_{0}}^2+4.79121 r {r_{0}}^3-5.8472 {r_{0}}^4) \\
                                & +p^6 (36.8094 r^4-69.4903 r^3 {r_{0}}+37.49 r^2 {r_{0}}^2-4.15303 r {r_{0}}^3+5.6248 {r_{0}}^4) \\
                                & +p^5 (-48.322 r^4+99.557 r^3 {r_{0}}-57.1551 r^2 {r_{0}}^2+0.840865 r {r_{0}}^3-0.70925 {r_{0}}^4) \\
                                & +p^4 (12.9333 r^4-21.1041 r^3 {r_{0}}+8.6512 r^2 {r_{0}}^2-0.0658137 r {r_{0}}^3-0.217336 {r_{0}}^4) \\
                                & +p^3 (0.768694 r^4-0.492765 r^3 {r_{0}}+1.99322 r^2 {r_{0}}^2+0.0218888 r {r_{0}}^3-0.0839553 {r_{0}}^4) \\
                                & +p^2 (-2.1015 r^4+1.63385 r^3 {r_{0}}-0.221489 r^2 {r_{0}}^2+0.0142061 r {r_{0}}^3-0.0117263 {r_{0}}^4) \\
                                & +p (-0.155143 r^4+0.0901895 r^3 {r_{0}}-0.0282534 r^2 {r_{0}}^2+0.000352495 r {r_{0}}^3-0.000442358 {r_{0}}^4) \\
                                & -0.005224 r^4+0.00352893 r^3 {r_{0}}-0.000651281 r^2 {r_{0}}^2,
 \end{aligned}
\end{equation}
        
\begin{equation}
\begin{aligned}
    \mathcal{D}^1_{\text{inv}} & = p^{11} r^2 (1. r^2-2. r {r_{0}}+1. {r_{0}}^2)+p^{10} r^2 (-33.5247 r^2+67.3011 r {r_{0}}-33.7764 {r_{0}}^2) \\
                               & +p^9 r^2 (139.634 r^2-287.439 r {r_{0}}+147.805 {r_{0}}^2)+p^8 r^2 (-211.343 r^2+449.189 r {r_{0}}-237.795 {r_{0}}^2) \\
                               & +p^7 r^2 (104.307 r^2-233.152 r {r_{0}}+127.187 {r_{0}}^2)+p^6 r^2 (36.8094 r^2-74.6521 r {r_{0}}+43.2038 {r_{0}}^2) \\
                               & +p^5 r^2 (-48.322 r^2+100.669 r {r_{0}}-57.9912 {r_{0}}^2)+p^4 r^2 (12.9333 r^2-20.9337 r {r_{0}}+8.67754 {r_{0}}^2) \\
                               & +p^3 r^2 (0.768694 r^2-0.696371 r {r_{0}}+1.94788 {r_{0}}^2)+p^2 r^2 (-2.1015 r^2+1.61854 r {r_{0}}-0.227976 {r_{0}}^2) \\
                               & +p r^2 (-0.155143 r^2+0.0896651 r {r_{0}}-0.0284235 {r_{0}}^2)+r^2 (-0.005224 r^2+0.00352893 r {r_{0}} \\
                               & -0.000651281 {r_{0}}^2),
\end{aligned}
\end{equation}

\begin{equation}
\begin{aligned}
    \mathcal{N}^2_{\text{inv}} & = p^9 (1. r^3-2.16356 r^2 {r_{0}}+1.01501 r {r_{0}}^2+0.148553 {r_{0}}^3)+p^8 (-8.63116 r^3+19.3261 r^2 {r_{0}} \\
                               & -9.37725 r {r_{0}}^2-1.31772 {r_{0}}^3)+p^7 (25.4073 r^3-61.4392 r^2 {r_{0}}+31.2754 r {r_{0}}^2+4.7492 {r_{0}}^3) \\
                               & +p^6 (-25.4485 r^3+80.1206 r^2 {r_{0}}-45.5637 r {r_{0}}^2-9.03094 {r_{0}}^3)+p^5 (-18.958 r^3-9.26834 r^2 {r_{0}} \\
                               & +18.1535 r {r_{0}}^2+9.73502 {r_{0}}^3)+p^4 (67.5176 r^3-88.9673 r^2 {r_{0}}+28.0939 r {r_{0}}^2-5.8473 {r_{0}}^3) \\
                               & +p^3 (-62.3959 r^3+97.8017 r^2 {r_{0}}-38.225 r {r_{0}}^2+1.71881 {r_{0}}^3)+p^2 (25.6081 r^3-41.9346 r^2 {r_{0}} \\
                               & +17.3586 r {r_{0}}^2-0.140684 {r_{0}}^3)+p (-4.42145 r^3+6.8935 r^2 {r_{0}}-2.85028 r {r_{0}}^2-0.0149354 {r_{0}}^3) \\
                               & +0.322118 r^3-0.368892 r^2 {r_{0}}+0.119768 r {r_{0}}^2,
\end{aligned}
\end{equation}

\begin{equation}
\begin{aligned}
    \mathcal{D}^2_{\text{inv}} & = p^9 r (1. r^2-2.16356 r {r_{0}}+1.16356 {r_{0}}^2)+p^8 r (-8.63116 r^2+19.3261 r {r_{0}}-10.695 {r_{0}}^2) \\
                               & +p^7 r (25.4073 r^2-61.4392 r {r_{0}}+36.0281 {r_{0}}^2)+p^6 r (-25.4485 r^2+80.1206 r {r_{0}}-54.6246 {r_{0}}^2) \\
                               & +p^5 r (-18.958 r^2-9.26834 r {r_{0}}+27.9908 {r_{0}}^2)+p^4 r (67.5176 r^2-88.9673 r {r_{0}}+22.0663 {r_{0}}^2) \\
                               & +p^3 r (-62.3959 r^2+97.8017 r {r_{0}}-36.3316 {r_{0}}^2)+p^2 r (25.6081 r^2-41.9346 r {r_{0}}+17.1297 {r_{0}}^2) \\
                               & +p r (-4.42145 r^2+6.8935 r {r_{0}}-2.84697 {r_{0}}^2)+r (0.322118 r^2-0.368892 r {r_{0}}+0.119768 {r_{0}}^2).
\end{aligned}
\end{equation}

\subsection{Logarithmic GB-Coupling Functional: \texorpdfstring{$f(\varphi)=\ln(\varphi)$}{TEXT}}

\begin{equation}
        g_{tt}(r) \approx \frac{\mathcal{N}^1_{log}}{\mathcal{D}^1_{log}}(1- \frac{r_0}{r}),
\end{equation}
\begin{equation}
         \sqrt{g_{tt}(r)g_{rr}(r)} \approx \frac{\mathcal{N}^2_{log}}{\mathcal{D}^2_{log}},
\end{equation}
    
 where   
\begin{equation}
 \begin{aligned}   
    \mathcal{N}^1_{\text{log}}  & = p^{13} (-0.0626084 r^3 {r_{0}}+0.0626084 r^2 {r_{0}}^2+0.0626084 r {r_{0}}^3-0.0626084 {r_{0}}^4) \\
                                & +p^{12} (1. r^4-0.782889 r^3 {r_{0}}-0.230599 r^2 {r_{0}}^2-1.21711 r {r_{0}}^3+1.2306 {r_{0}}^4) \\
                                & +p^{11} (-18.2212 r^4+29.3638 r^3 {r_{0}}-10.8868 r^2 {r_{0}}^2+7.28478 r {r_{0}}^3-7.54055 {r_{0}}^4) \\
                                & +p^{10} (96.59 r^4-177.29 r^3 {r_{0}}+79.5585 r^2 {r_{0}}^2-19.3886 r {r_{0}}^3+20.5203 {r_{0}}^4) \\
                                & +p^9 (-234.614 r^4+457.171 r^3 {r_{0}}-221.897 r^2 {r_{0}}^2+23.902 r {r_{0}}^3-24.2931 {r_{0}}^4) \\
                                & +p^8 (282.436 r^4-551.051 r^3 {r_{0}}+275.112 r^2 {r_{0}}^2-7.20824 r {r_{0}}^3-1.91936 {r_{0}}^4) \\
                                & +p^7 (-133.991 r^4+164.055 r^3 {r_{0}}-46.0945 r^2 {r_{0}}^2-13.4629 r {r_{0}}^3+41.7195 {r_{0}}^4) \\
                                & +p^6 (-32.4983 r^4+341.82 r^3 {r_{0}}-300.797 r^2 {r_{0}}^2+12.9375 r {r_{0}}^3-53.4003 {r_{0}}^4) \\
                                & +p^5 (34.5271 r^4-420.143 r^3 {r_{0}}+404.167 r^2 {r_{0}}^2+0.0115351 r {r_{0}}^3+31.9943 {r_{0}}^4) \\
                                & +p^4 (28.0229 r^4+178.063 r^3 {r_{0}}-241.727 r^2 {r_{0}}^2-4.86976 r {r_{0}}^3-8.90541 {r_{0}}^4) \\
                                & +p^3 (-30.9869 r^4-12.6459 r^3 {r_{0}}+69.661 r^2 {r_{0}}^2+2.21647 r {r_{0}}^3+0.507243 {r_{0}}^4) \\
                                & +p^2 (8.38109 r^4-10.7164 r^3 {r_{0}}-6.17049 r^2 {r_{0}}^2-0.251471 r {r_{0}}^3+0.12598 {r_{0}}^4) \\
                                & +p (-1.03581 r^4+2.49996 r^3 {r_{0}}-0.826428 r^2 {r_{0}}^2-0.0167726 r {r_{0}}^3+0.0233143 {r_{0}}^4) \\
                                & +0.389358 r^4-0.281216 r^3 {r_{0}}+0.0686144 r^2 {r_{0}}^2,
 \end{aligned}                               
\end{equation}
      
\begin{equation}
\begin{aligned}
    \mathcal{D}^1_{\text{log}} & = p^{12} r^2 (1. r^2-2. r {r_{0}}+1. {r_{0}}^2)+p^{11} r^2 (-18.2212 r^2+36.6579 r {r_{0}}-18.4367 {r_{0}}^2) \\
                               & +p^{10} r^2 (96.59 r^2-197.003 r {r_{0}}+100.413 {r_{0}}^2)+p^9 r^2 (-234.614 r^2+483.3 r {r_{0}}-248.622 {r_{0}}^2) \\
                               & +p^8 r^2 (282.436 r^2-565.092 r {r_{0}}+281.412 {r_{0}}^2)+p^7 r^2 (-133.991 r^2+161.605 r {r_{0}}-19.988 {r_{0}}^2) \\
                               & +p^6 r^2 (-32.4983 r^2+345.678 r {r_{0}}-336.459 {r_{0}}^2)+p^5 r^2 (34.5271 r^2-417.829 r {r_{0}}+424.172 {r_{0}}^2) \\
                               & +p^4 r^2 (28.0229 r^2+175.038 r {r_{0}}-246.175 {r_{0}}^2)+p^3 r^2 (-30.9869 r^2-11.7993 r {r_{0}}+69.4436 {r_{0}}^2) \\
                               & +p^2 r^2 (8.38109 r^2-10.8149 r {r_{0}}-6.01301 {r_{0}}^2)+p r^2 (-1.03581 r^2+2.54012 r {r_{0}}-0.815274 {r_{0}}^2) \\
                               & +r^2 (0.389358 r^2-0.281216 r {r_{0}}+0.0686144 {r_{0}}^2),
\end{aligned}
\end{equation}

\begin{equation}
\begin{aligned}
    \mathcal{N}^2_{\text{log}} & = p^9 (1. r^3-1.5934 r^2 {r_{0}}+0.9878 r {r_{0}}^2-0.394399 {r_{0}}^3)+p^8 (-11.4341 r^3+18.4084 r^2 {r_{0}} \\
                               & -9.70115 r {r_{0}}^2+2.72685 {r_{0}}^3)+p^7 (52.5869 r^3-85.6446 r^2 {r_{0}}+40.9887 r {r_{0}}^2-7.89466 {r_{0}}^3) \\
                               & +p^6 (-129.875 r^3+213.791 r^2 {r_{0}}-96.5185 r {r_{0}}^2+12.2873 {r_{0}}^3)+p^5 (190.394 r^3-316.075 r^2 {r_{0}} \\
                               & +137.73 r {r_{0}}^2-10.895 {r_{0}}^3)+p^4 (-170.683 r^3+284.439 r^2 {r_{0}}-121.325 r {r_{0}}^2+5.25315 {r_{0}}^3) \\
                               & +p^3 (92.159 r^3-152.461 r^2 {r_{0}}+64.1216 r {r_{0}}^2-1.0636 {r_{0}}^3)+p^2 (-28.3648 r^3+45.1052 r^2 {r_{0}} \\
                               & -18.6363 r {r_{0}}^2-0.0535942 {r_{0}}^3)+p (4.63158 r^3-6.37023 r^2 {r_{0}}+2.46462 r {r_{0}}^2+0.0339455 {r_{0}}^3) \\
                               &-0.413973 r^3+0.400205 r^2 {r_{0}}-0.112264 r {r_{0}}^2,
\end{aligned}
\end{equation}

\begin{equation}
\begin{aligned}
    \mathcal{D}^2_{\text{log}} & = p^9 r (1. r^2-1.5934 r {r_{0}}+0.593401 {r_{0}}^2)+p^8 r (-11.4341 r^2+18.4084 r {r_{0}}-6.9743 {r_{0}}^2) \\
                               & +p^7 r (52.5869 r^2-85.6446 r {r_{0}}+33.0761 {r_{0}}^2)+p^6 r (-129.875 r^2+213.791 r {r_{0}}-84.1127 {r_{0}}^2) \\
                               & +p^5 r (190.394 r^2-316.075 r {r_{0}}+126.509 {r_{0}}^2)+p^4 r (-170.683 r^2+284.439 r {r_{0}}-115.594 {r_{0}}^2) \\ 
                               & +p^3 r (92.159 r^2-152.461 r {r_{0}}+62.6652 {r_{0}}^2)+p^2 r (-28.3648 r^2+45.1052 r {r_{0}}-18.518 {r_{0}}^2) \\
                               & +p r (4.63158 r^2-6.37023 r {r_{0}}+2.46727 {r_{0}}^2)+r (-0.413973 r^2+0.400205 r {r_{0}}-0.112264 {r_{0}}^2).
\end{aligned}
\end{equation}

\end{appendices}

\noindent\hrulefill

\printbibliography
\end{document}